\documentclass[prd,groupedaddress,showpacs,nofootinbib]{revtex4}
\usepackage{graphicx}
\usepackage{dcolumn}
\usepackage{bm}
\usepackage{amssymb}
\usepackage{amsmath}
\usepackage{epsfig}
\usepackage[dvips]{color}
\usepackage{hhline}
\usepackage{color}

\def\lapp{\mathrel{\rlap{\raise.5ex\hbox{$<$}}
                    {\lower.5ex\hbox{$\sim$}}}}
\def\gapp{\mathrel{\rlap{\raise.5ex\hbox{$>$}}
                    {\lower.5ex\hbox{$\sim$}}}}

{\newcommand{\lsim}{\mbox{\raisebox{-.6ex}{~$\stackrel{<}{\sim}$~}}}
{\newcommand{\gsim}{\mbox{\raisebox{-.6ex}{~$\stackrel{>}{\sim}$~}}}

\newcommand{\bmt}{\begin{pmatrix}}
\newcommand{\emt}{\end{pmatrix}}
\newcommand{\ba}{\begin{array}{c}}
\newcommand{\ea}{\end{array}}
\newcommand{\be}{\begin{equation}}
\newcommand{\ee}{\end{equation}}
\newcommand{\bea}{\begin{eqnarray}}
\newcommand{\eea}{\end{eqnarray}}

\newcommand{\bi}{\begin{itemize}}
\newcommand{\ei}{\end{itemize}}

\newcommand{\baz}{\begin{array}{cc}}

\newcommand{\mathsym}[1]{{}}

\newcommand{\bt}{\begin{tabular}}
\newcommand{\et}{\end{tabular}}

\newcommand{\benu}{\begin{enumerate}}
\newcommand{\eenu}{\end{enumerate}}

\newcommand{\bav}{\begin{array}{cccc}}

\begin{document}
\title{\bf Double Beta Decay, Lepton Flavour Violation and Collider Signatures\\[0.5mm]of Left-Right 
Symmetric Models with Spontaneous $D$ Parity Breaking} 
\author{Frank F. Deppisch}
\email{f.deppisch@ucl.ac.uk}
\author{Tomas E. Gonzalo}
\email{tomas.gonzalo.11@ucl.ac.uk}
\affiliation{Department of Physics and Astronomy, University College London, London, WC1E 6BT, United Kingdom}
\author{Sudhanwa Patra}
\email{sudha.astro@gmail.com}
\affiliation{Centre of Excellence in Theoretical and Mathematical Sciences, Siksha \textquoteleft O\textquoteright 
Anusandhan University, Bhubaneswar, 751030, India}
\author{Narendra Sahu}
\email{nsahu@iith.ac.in}
\affiliation{Department of Physics, Indian Institute of Technology, Hyderabad, Yeddumailaram, 502205, 
Telengana, India}
\author{Utpal Sarkar}
\email{utpal@prl.res.in}
\affiliation{Physical Research Laboratory, Ahmedabad, 380009, Gujrat, India}

\begin{abstract}
We propose a class of left-right symmetric models (LRSMs) with spontaneous $D$ parity breaking, where $SU(2)_R$ breaks at the TeV scale while discrete left-right symmetry breaks around $10^{9}$~GeV. By embedding this framework in a non-supersymmetric $SO(10)$ Grand Unified Theory (GUT) with Pati-Salam symmetry as the highest intermediate breaking step, we obtain $g_R / g_L \approx 0.6$ between the right- and left-handed gauge couplings at the TeV scale. This leads to a suppression of beyond the Standard Model phenomena induced by the right-handed gauge coupling. Here we focus specifically on the consequences for neutrinoless double beta decay, low energy lepton flavour violation and LHC signatures due to the suppressed right handed currents. Interestingly, the reduced $g_R$ allows us to interpret an excess of events observed recently in the range of 1.9 TeV to 2.4 TeV by the CMS group at the LHC as the signature of a right handed gauge boson in LRSMs with spontaneous $D$ parity breaking. Moreover, the 
reduced right-handed gauge coupling also strongly suppresses the non-standard contribution of heavy states to the neutrinoless double beta decay rate as well as the amplitude of low energy lepton flavour violating processes. In a dominant type-II Seesaw mechanism of neutrino mass generation, we find that both sets of observables provide stringent and complimentary bounds which make it challenging to observe the scenario at the LHC.  

\end{abstract}

\pacs{98.80.Cq,14.60.Pq}
\maketitle

\newpage
\section{Introduction}

The current low energy data from solar, atmospheric and reactor neutrinos \cite{oscillation_expts} established the oscillation hypothesis with very small masses ($\leq 1$ eV) for the three generations of light neutrinos. Depending on whether neutrinos are Dirac~\cite{Dirac:1925} (having distinct antiparticles) or Majorana~\cite{Majorana:1937vz} (they are their own antiparticles) fermions, these masses originate from the corresponding Dirac or Majorana mass terms. The goal of the current experimental neutrino programme is to determine the nine degrees of freedom of the neutrino sector: three light neutrino masses ($m_1, m_2, m_3$), three mixing angles ($\theta_{12}, \theta_{23}, \theta_{13}$) and potentially up to three CP phases: one Dirac phase ($\delta$) and two Majorana phases ($\alpha, \beta$). The oscillation experiments allowed us so far to measure the two mass squared differences and the three mixing angles, but we are yet to determine the absolute mass scale, the presence of $CP$ violation and 
whether neutrinos are Dirac or Majorana fermions.

If neutrinos are Dirac fermions then lepton number is an exact symmetry of the low energy effective theory. On the other hand, if neutrinos are Majorana fermions, there will be a violation of lepton number by two units. The latter would necessarily predict neutrinoless double beta decay ($0\nu\beta\beta$), i.e. the nuclear decay $(A,Z) \to (A, Z+2) + 2e^-$ of various nuclei. The experimental non-observation currently provides lower bounds on the half life of this process in various isotopes, of the order $T_{1/2} \gsim 2 \times 10^{25}$~yr~\cite{Klapdor-bound,Klapdor-claim,gerda,Igex,kamland}. This can be translated to a bound on the so-called effective $0\nu\beta\beta$ mass parameter as $m_{ee} \leq 0.2 - 0.6$~eV, with a large uncertainty due to the theoretical error on the relevant nuclear matrix elements. Future experiments aim to improve the sensitivity on the $0\nu\beta\beta$ by about an order of magnitude, with a corresponding improvement in $m_{ee}$ by a factor of 3. Thus the Majorana nature of light neutrinos will be probed at future $0\nu\beta\beta$ experiments which will not only shed light on the absolute mass scale of the left-handed (LH) neutrinos but may also indicate the mass hierarchy and mass mechanism for LH active neutrinos (for details, see ref.~\cite{review1}). If we assume that the SM light Majorana neutrinos are only contributing to this rare $0\nu\beta\beta$ decay, then the present experimental bound on the $0\nu\beta\beta$ half life can be saturated with a quasi-degenerate (QD) pattern of light neutrinos, while normal hierarchy (NH) and inverted hierarchy (IH) patterns of light neutrinos remain unreachable within the sensitivities of current experiments. 
On the other hand, Planck~\cite{planck} and other astrophysical observations give a stringent bound on sum of masses of the light neutrinos, i.e, $\sum_i m_i \leq 0.23$~eV (95\% C.L.) which is in tension with the QD nature of the neutrinos and hence does not support the idea of light neutrinos being QD which saturate the present $0\nu\beta\beta$ experimental bound. Hence, if this rare decay process were to be observed with currently running experiments, it would indicate new physics contributions to $0\nu\beta\beta$ decay. 

At present various seesaw mechanisms exist which could elegantly explain the small Majorana masses of three active neutrinos without fine tuning. In the type-I seesaw mechanism~\cite{type-I-group} three right-handed (RH) neutrinos, which are singlets under the $SU(2)_L$ gauge group, are added to the Standard Model (SM). Integrating out the RH neutrinos with heavy Majorana masses of the order $M_R \approx 10^{14}$~GeV, the light neutrinos masses are generated as $m_\nu \approx M_\text{EW}^2 / M_R$. Generally, the Majorana masses of the singlet RH neutrinos are free parameters of the model and hence can vary from the GUT scale down to TeV scale or even lower. On the other hand, in the type-II seesaw mechanism~\cite{type-II-group} one adds a scalar triplet $\Delta$ with hyper charge 2 to the SM spectrum. After electroweak phase transition, $\Delta$ acquires an induced vacuum expectation value (VEV) and generates a Majorana mass matrix $m_\nu = f \langle \Delta \rangle $ for the three active neutrinos through its symmetric coupling $f \Delta L L$ to the lepton doublet in the SM. Note that the masses of the RH neutrinos and the scalar triplet are not controlled by the SM gauge group. 

A well motivated framework of beyond the Standard Model physics is the left-right symmetric model (LRSM) which is based on the gauge group $SU(2)_L \times SU(2)_R \times U(1)_{\rm B-L}$~\cite{LR}. In this case the masses of RH neutrinos and scalar triplets are governed by the scale of $SU(2)_R \times U(1)_{\rm B-L}$ breaking. The neutrino mass matrix receives contributions from both type-I and type-II seesaw mechanisms. If the breaking scale of $SU(2)_R \times U(1)_{\rm B-L}$ is at the TeV scale, the RH neutrinos, the scalar triplet and the RH gauge bosons acquire TeV scale masses. This leads to many interesting phenomena in the low energy effective theory. In particular, here we will focus on $0\nu\beta\beta$, lepton flavor violation (LFV) and collider signatures of these TeV scale particles. 

There are many studies~\cite{ibarra, MSV, pascoli, Pascoli:2013fiz, 0nubb-rnp-1980, 0nubb-vergados, 0nubb-hirsch, 0nubb-tello-2011, 0nubb-chakra-2012, 0nubb-frank-2012, 0nubb-psb-lee, 0nubb-rod-barry, 0nubb-rod-ilc, 0nubb-senj-2013, 0nubb-spatra-plb, 0nubb-spatra-jhep, 0nubb-Nemevsek, 0nubb-spatra-prd, 0nubb-huang, 0nubb-sruba-2014} of various TeV scale models and their phenomenological consequences. With respect to neutrinoless double beta decay, LRSMs can generate a large number of different non-standard contributions to $0\nu\beta\beta$ involving purely LH and RH currents as well as diagrams involving both LH and RH currents and a heavy triplet Higgs. In combination with the light neutrino masses arising in the corresponding seesaw mechanisms, this provides stringent bounds on LRSMs. In addition, analyses made over the recent years explore the correlation between $0\nu\beta\beta$, LFV and collider signatures in certain TeV scale LRSMs involving purely RH currents via heavy neutrino plus Higgs triplet exchange by assuming either a purely type-II seesaw dominance or a type-I and/or type-II seesaw dominance, see for example \cite{0nubb-tello-2011} and \cite{0nubb-chakra-2012, 0nubb-rod-barry}, respectively. The studies of $0\nu\beta\beta$ decays in LRSMs put constraints on the heavy RH gauge bosons and RH neutrino masses which have to be compatible with the direct search limits from accelerator experiments like the LHC~\cite{LHC-RR} and from low energy LFV searches. All these studies so far assumed an explicitly symmetric structure of the left-right model at TeV scales, i.e $g_L = g_R$. Although these models provide a rich phenomenology while keeping a low scale of left-right symmetry breaking, it is difficult to justify them while being consistent with gauge coupling unification in a non-supersymmetric framework.

However, there exists another class of LRSMs with spontaneous $D$ parity breaking~\cite{Dparity, Dparity-1, patra-dparity} where a discrete left-right symmetry called $D$ parity is broken at a higher scale compared to the $SU(2)_R$ symmetry breaking scale. As a result, an asymmetry is generated between the left- and RH Higgs fields making the coupling constants of $SU(2)_R$ and $SU(2)_L$ evolve separately under the renormalization group from the scale of $D$ parity breaking down to the TeV scale where the $SU(2)_R$ gauge symmetry is allowed to break. Consequently, the corresponding gauge couplings strengths are no longer equal, $g_L \neq g_R$ at the TeV scale which crucially affects low energy and LHC processes. Hence the effect of $g_{L} \neq g_{R}$ should be examined carefully while deriving important conclusions at TeV scales.

In this paper we make an attempt to study the effect of $g_{L} \neq g_{R}$ in $0\nu\beta\beta$ decay, LFV and collider processes involving RH currents in a class of TeV scale LRSMs with spontaneous $D$ parity breaking. We ensure that the masses of RH particles are of the order of the TeV scale by extending the left-right gauge group $SU(2)_L\times SU(2)_R \times U(1)_{\rm B-L}$ with a $D$ parity and make sure that the discrepancy between the $SU(2)_L$ coupling $g_{L}$ and the $SU(2)_R$ coupling $g_{R}$ is indeed sufficiently large. By embedding this framework in a non-supersymmetric $SO(10)$ Grand Unified Theory (GUT) with Pati-Salam symmetry at the highest intermediate breaking step, we obtain $g_R / g_L \approx 0.6$ at th TeV scale. Below the GUT scale, the $D$ parity breaks first at a high scale $\approx 10^9$~GeV below which $g_L$ and $g_R$ evolve differently. Moreover, the breaking creates a large mass splitting between the LH and RH scalar particles. We assume that the LH scalar particles are heavy leaving the RH scalar particles at TeV scales. Subsequently, $SU(2)_R$ breaks to $U(1)_R$ at a scale of $\approx 10$~TeV, and the RH $W_R$ boson acquires a TeV scale mass. In the next step, $U(1)_R \times U(1)_{\rm B-L}$ breaks at a scale of ${\cal O}(\rm TeV)$, leading to RH $Z_R$ boson and neutrino masses potentially accessible at the LHC. Consequently, the heavy RH states can all be as light as the TeV scale. Moreover, the suppressed gauge coupling $g_R$ allows us to interpret an excess of events observed in the range of 1.9 TeV to 2.4 TeV by the CMS group~\cite{cms_excess} at LHC as the signature of a right handed gauge boson of LRSMs with spontaneous $D$ parity breaking as pointed out in \cite{Deppisch:2014qpa} and in subsequent works~\cite{Heikinheimo:2014tba, Aguilar-Saavedra:2014ola}. 

The plan of the paper is sketched as follows. In Section~\ref{eq:lrsm-dparity}, we briefly outline the TeV scale LRSM invoked with spontaneous $D$ parity breaking, embed the framework of a TeV scale LRSM in a non-supersymmetric $SO(10)$ model with Pati-Salam symmetry at the highest intermediate breaking step and discuss the one-loop renormalization group evolution for gauge coupling unification. The neutrino mass generation of three active light neutrinos via the dominant type-II seesaw mechanism and the relation between light and heavy neutrinos are presented here as well. In Section~\ref{sec:0nu2beta}, we analyze the possible contributions to $0\nu\beta\beta$ decay within our left-right model, i.e. with $g_L \neq g_R$ and with special emphasis on $W_R-W_R$ mediated diagrams. We then discuss the possible flavour violating effects and collider signatures of this particular TeV scale left-right model in sections~\ref{sec:LFV} and \ref{sec:collider}, respectively. Finally, we summarize our results and conclude in Section~\ref{sec:conclusions}.

\section{Left-Right Symmetric Models with Spontaneous $D$ Parity Breaking}
\label{eq:lrsm-dparity}
The purpose of this section is two fold: i) firstly, to briefly discuss left-right symmetric model with spontaneous $D$ parity breaking ($LRSM_{D \hspace*{-0.19cm}/}$) while keeping $W_R$, $Z_R$ gauge boson masses around the TeV scale in order to have dominant non-standard contributions for $0\nu\beta\beta$ decay, lepton flavor violation and associated collider signatures, ii) secondly, to provide type II seesaw dominance for light neutrino mass generation mechanism and yield direct relation between light and heavy neutrino mass eigenvalues so that one can easily deduce the complementary relation between $0\nu\beta\beta$, LFV and Collider processes including new physics contributions. 

The basic gauge group is $SU(2)_L \times SU(2)_R \times U(1)_{B-L} \times D$, where $D$ denotes the discrete left-right symmetry or $D$ parity. The matter sector of the model includes leptons and quarks transforming under the left-right symmetric group as
\begin{align}
	\ell_{L} &=
		\begin{pmatrix}
		  \nu_{L}\\
	    e_{L}
	  \end{pmatrix} \equiv(2,1,-1), \quad
	\ell_{R} =
	\begin{pmatrix} 
		\nu_{R}\\
	  e_{R}
	\end{pmatrix} \equiv(1,2,-1), \quad \nonumber \\
	Q_{L} &=
	  \begin{pmatrix}
	    u_{L} \\
	    d_{L}
	  \end{pmatrix} \equiv(2,1,{\frac{1}{3}}), \quad
	Q_{R} =
	  \begin{pmatrix}
	    u_{R}\\
	    d_{R}
	  \end{pmatrix} \equiv(1,2,{\frac{1}{3}}).
\end{align}
It is quite clear that the RH charged lepton of each family which was an isospin singlet under the SM gauge group gets a new partner $\nu_R$. The two form an isospin doublet under $SU(2)_R$ of the left-right symmetric gauge group $SU(2)_L \times SU(2)_R \times U(1)_{B-L}\times D$. Similarly, in the quark-sector, the right handed up and down quarks of each family, which were isospin singlets under the SM gauge group, combine to form the isospin doublet under $SU(2)_R$. As a result and before the left-right symmetry breaking, both left- and right-handed leptons and quarks enjoy equal strength of interactions. This explains that parity is a good quantum number in the LRSM in contrast to the SM where the LH particles are preferential under the electroweak interaction. 

To implement the symmetry breaking, the Higgs sector of the present model consists of a $SU(2)$ singlet scalar field $\sigma$ which is odd under the discrete $D$ parity, two $SU(2)_L$ triplets $\Delta_L$ and $\Omega_L$, two $SU(2)_R$ triplets $\Delta_R$, $\Omega_R$ and a bidoublet $\Phi$ which contains two copies of the SM Higgs. Under $SU(2)_L\times SU(2)_R \times U(1)_{B-L}$, the quantum numbers of the these Higgs fields are given as
\begin{align}
	\Delta_{L} &= 
	\begin{pmatrix}
		\delta_{L}^{+}/\sqrt{2} & \delta_{L}^{++}\\
		\delta_{L}^{0} & -\delta_{L}^{+}/\sqrt{2}
	\end{pmatrix} \equiv
	(3,1,-2), & 
	\Delta_{R} &= 
	\begin{pmatrix}
		\delta_{R}^{+}/\sqrt{2} & \delta_{R}^{++}\\
		\delta_{R}^{0} & -\delta_{R}^{+}/\sqrt{2}
	\end{pmatrix} \equiv
	(1,3,-2), \nonumber \\
	\Omega_{L} &= 
	\begin{pmatrix}
		\omega_{L}^{0} & \omega_{L}^{+}/\sqrt{2}\\
		\omega_{L}^{-}/\sqrt{2} & -\omega_{L}^{0}
	\end{pmatrix} \equiv (3,1,0), &
	\Omega_{R} &=
	\begin{pmatrix}
		\omega_{R}^{0} & \omega_{R}^{+}/\sqrt{2}\\
		\omega_{R}^{-}/\sqrt{2} & -\omega_{R}^{0}
	\end{pmatrix} \equiv (1,3,0), \\
	\Phi &=
	\begin{pmatrix}
		\phi_{1}^{0} & \phi_{1}^{+}\\
		\phi_{2}^{-} & \phi_{2}^{0}
	\end{pmatrix} \equiv
	(2,2,0), &
	\sigma &\equiv (1,1,0). \nonumber 
\end{align}
The Higgs scalars and vector bosons of the present model transform under the operation of $D$ parity as
\begin{align}
	\psi_{L,R} \longrightarrow \psi_{R,L}, \quad
	\Phi \longrightarrow \Phi^T, \quad
	\Delta_{L,R} \longrightarrow \Delta_{R,L}, \quad
	\Omega_{L,R} \longrightarrow  \Omega_{R,L}, \quad
	\sigma \longrightarrow -\sigma, \quad
	W_{L,R} \longrightarrow W_{R,L}.
\label{p-parity}
\end{align}
The VEVs assigned for different Higgs scalars are given below
\begin{align}
	\langle \Phi \rangle &=\frac{1}{\sqrt{2}}
	\begin{pmatrix}
		v_u & 0  \\
		0   & v_d
	\end{pmatrix}\, , \quad 
	\langle \Delta_{L} \rangle = \frac{1}{\sqrt{2}}
	\begin{pmatrix}
		0     & 0\\
		v_L   & 0
	\end{pmatrix}\, , \quad 
	\langle \Delta_{R} \rangle = \frac{1}{\sqrt{2}}
	\begin{pmatrix}
		0     & 0\\
		v_R   & 0
	\end{pmatrix}\, , \nonumber \\
	\langle \Omega_{L} \rangle &= \frac{1}{\sqrt{2}}
	\begin{pmatrix}
		\omega_L & 0\\
		0        & \omega_L
	\end{pmatrix}\, , \quad 
	\langle \Omega_{R} \rangle = \frac{1}{\sqrt{2}}
	\begin{pmatrix}
		\omega_R & 0\\
		0        & \omega_R
	\end{pmatrix}\, , \quad \langle \sigma \rangle = M_P \, .
\end{align}
The LRSM gauge group breaks as $SU(2)_L \times SU(2)_R \times U(1)_{B-L} \times D 
\stackrel{\langle \sigma \rangle}{\longrightarrow} SU(2)_L 
\times SU(2)_R \times U(1)_{B-L}$ since the discrete left-right symmetry ($D$ parity) breaks spontaneously after the scalar field $\sigma$, odd under $D$ parity, is assigned a VEV without breaking $SU(2)_R$. The subsequent step of symmetry breaking from $SU(2)_L \times SU(2)_R \times U(1)_{B-L}$ to the standard model gauge group $SU(2)_L \times U(1)_Y$ is done via two steps: firstly, the $SU(2)_R$ breaks down to $U(1)_R$ without breaking rank of the group by giving a VEV to the RH Higgs triplet $\Omega_R$ with zero $(B-L)$ charge at scale $M_\Omega$ and secondly, $U(1)_R \times U(1)_{B-L}$ breaks down to $U(1)_Y$ by assigning a VEV to $\langle\Delta^0_R(1,1,-2,1) \rangle\sim v_R/\sqrt{2}$ at a latter stage $M_{B-L}$. The SM gauge group $SU(2)_L \times U(1)_Y$ breaks to $U(1)_{\rm em}$ by giving a VEV to the SM Higgs doublet contained in the bidoublet $\Phi$. After spontaneous symmetry breaking, the RH charged gauge boson $W_R$ gets its mass from $\Omega_R$ VEV at around $(6-10)$ TeV while the extra neutral RH gauge boson $Z_R$ obtained its mass from VEV of $\Delta_R$ around $4-5$ TeV. Thus, it is clear that the origin of mass of $W_R$ completely decouples from $Z_R$ mass origin and hence, the Collider studies of these extra gauge bosons and the corresponding mass bounds should be revived again. 

After some simple algebra, one can derive the analytic expressions for various Higgs scalar masses as well as gauge boson masses 
as 
\begin{align}
	M^2_{W_R}      &\approx  \frac{g^2_{R} \omega^2_R}{2}, & 
	M^2_{W_L}      &\approx \frac{g^2_{L} v^2}{4}, \nonumber \\
	M^2_{Z_R}      &\approx \frac{1}{2} \left(g^2_{B-L}+g^2_{R} \right) (v^2 + 4 v^2_R), &
	M^2_{Z_L}      &\approx \frac{M^2_{W_L}}{\cos^2 \theta_W}, \\
	M^2_{\Delta_R} &\approx \mu^2_{\Delta_R} 
	                      - \lambda \langle \sigma\rangle M, &
	M^2_{\Delta_L} &\approx \mu^2_{\Delta_L} 
	                      + \lambda \langle \sigma\rangle M, \nonumber\\
	M^2_{\Omega_R} &\approx \mu^2_{\Omega_R} 
	                      - \lambda^\prime \langle \sigma\rangle M^\prime, &
	M^2_{\Omega_L} &\approx \mu^2_{\Omega_L} 
	                      + \lambda^\prime \langle \sigma\rangle M^\prime, \nonumber
\end{align}
where $v=\sqrt{v_u^2 + v_d^2}=\mbox{246\,GeV}$, is the electroweak scale, $\theta_W$ is the Weinberg mixing angle such that $\sin^2 \theta_W \simeq 0.23116$. The other parameters like $\lambda$ and $\lambda^\prime$ are the trilinear Higgs coupling and $\langle \sigma \rangle$, $M$, $M'$ are of the order of $D$ parity breaking scale. The LH fields i.e, $\Delta_L$ and $\Omega_L$ remain heavy when $SU(2)_R$ gauge symmetry breaking scale decouples completely from the $D$ parity breaking scale~\cite{Dparity}. As a result, the low energy Lagrangian has an invariance under Left-Right gauge group but not under $D$ parity.

\subsection{$SO(10)$ GUT Embedding}
\label{sec:GUT}
We note that the gauge coupling $g_R$ corresponding to the gauge group $SU(2)_R$ is a free parameter within the low energy asymmetric left-right model due to spontaneous $D$ parity breaking. However, we can calculate it by embedding the left-right gauge group in a non-supersymmetric $SO(10)$ GUT and examine the renormalization group (RG) evolution of the gauge couplings at the one-loop level. The gauge coupling unification then predicts the value of $g_R$ at TeV scales. 

Here we make an attempt to embed the model in a non-SUSY $SO(10)$ GUT by considering the Pati-Salam symmetry at the highest intermediate breaking step (see \cite{0nubb-spatra-jhep, PS-dparity} for details) as
\begin{equation}
\label{eq:BreakingChain}
	SO(10) \mathop{\longrightarrow}^{M_U}_{}      \mathcal{G}_{224D} \,
  	     \mathop{\longrightarrow}^{M_P}_{}      \mathcal{G}_{224}  \,
   	     \mathop{\longrightarrow}^{M_C}_{}      \mathcal{G}_{2213} \,
   	     \mathop{\longrightarrow}^{M_\Omega}_{} \mathcal{G}_{2113} \,
   	     \mathop{\longrightarrow}^{M_{B-L}}_{}  \mathcal{G}_{SM}   \,
   	     \mathop{\longrightarrow}^{M_{Z}}_{}    \mathcal{G}_{13}. 
\end{equation}
The breaking of $SO(10)$ to the Pati-Salam group and $D$ parity invariance, $SO(10) \rightarrow G_{224D}$, can be achieved by giving a VEV to $G_{224}$ singlets residing in a $\{54\}_H$-plet Higgs of $SO(10)$ which is even under $D$ parity. The $\mathcal{G}_{224}$ multiplet $(1,1,1) \subset \{210\}_H$, being odd under $D$ parity, is responsible for the second stage of symmetry breaking: $G_{224D} \rightarrow G_{224}$. We denote this scale by $M_P$ where $D$ parity invariance breaks. This stage is crucial as  it results in different masses for the fields $\Delta_L$, $\Omega_L$ and $\Delta_R$, $\Omega_R$, and therefore the $SU(2)_L$ and $SU(2)_R$ gauge couplings evolve differently from this scale downwards, yielding $g_L \neq g_R$. The key role of the Pati-Salam symmetry with or without its embedding in the $SO(10)$ model is to give the ratio $g_R / g_L$ a drastically different value from unity.

The next stage of symmetry breaking $G_{224} \rightarrow G_{2213}$ is achieved by assigning a VEV to Pati-Salam multiplet $(1,1,15) \subset \{210\}_H$ where $SU(4)_C$ breaks to $U(1)_{B-L} \times SU(3)_C$ at a scale $M_C$. The breaking of Pati-Salam symmetry occurs around $10^5-10^6$~GeV and thus, the di-quark Higgs scalars and leptoquark gauge bosons get their masses around the same scale. The model has potential to predict neutron-antineutron oscillation via the exchange of di-quark Higgs scalars with a mixing time $\tau_{n-\overline{n}}$ with mixing time very close to limits derived by recent ongoing search experiments \cite{n-nbar}.

The subsequent stage of symmetry breaking $G_{2213} \rightarrow G_{2113}$, where $SU(2)_R$ breaks to $U(1)_R$ without breaking the rank of the group, is generated by the VEV of the $\mathcal{G}_{2213}$ component under the Pati-Salam gauge group, $\Omega_R(1, 3, 0, 1) \subset (1, 3, 15) \subset \{210\}_H$. The corresponding scale is denoted as $M_\Omega$ and which is of the order 6-10~TeV. As a result we get the RH charged gauged boson mass $M_{W_R}$ around 2-3~TeV. Subsequently, the RH triplet Higgs field $\Delta_R \subset \Delta_R (1,3,-2,1)$ belonging to the Higgs representation $126_H$ acquires a VEV and thus breaking $G_{2113}\rightarrow G_{SM}$, where $U(1)_R \times U(1)_{B-L}$ breaks to $U(1)_Y$. This is another important stage of symmetry breaking which results in TeV scale masses for extra neutral gauge boson $Z_R$, the RH neutrinos and RH Higgs triplets. The final stage of symmetry breaking $G_{SM} \rightarrow SU(3)_C \times U(1)_{\rm em}$ occurs via the VEV of the SM doublet $\phi \subset 10_H$.

\begin{figure}[t]
\centering
\includegraphics[width=0.6\columnwidth]{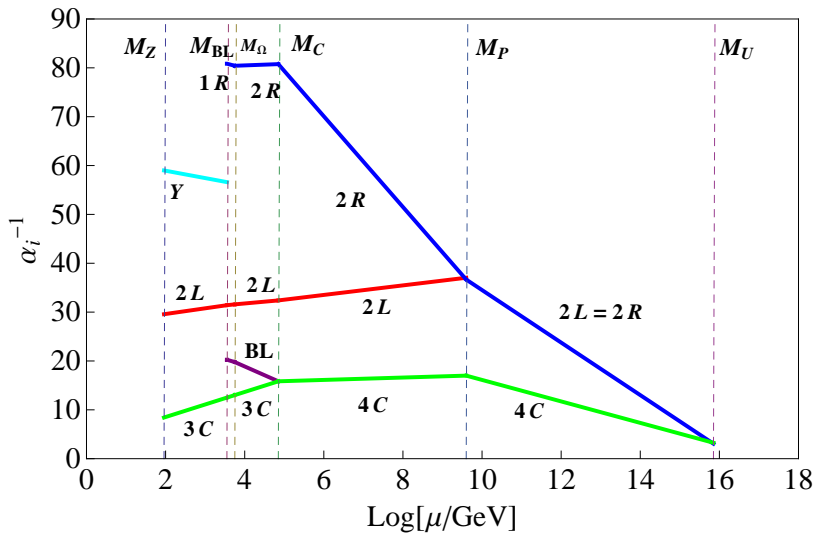}
\caption{RG evolution of gauge couplings yielding $g_R/g_L=0.6 $ within a non-SUSY $SO(10)$ GUT, where $\mathcal{G}_{224}$ and $\mathcal{G}_{2213}$ occur as intermediate symmetry breaking steps. The plot shows the running of $\alpha_i^{-1} = 4\pi/g_i^2$ for the couplings $g_i$ of the relevant gauge groups as a function of the energy scale $\mu$ where $i = 2L, 2R, 4C, 1R, BL, 3C, Y$. The gauge breaking scales, as denoted by the dashed vertical lines, consistent with gauge coupling unification are: $M_Z = 91.2$~GeV, $M_{B-L} \approx 4$~TeV, $M_{\Omega} \approx 6$~TeV, $M_C \approx 10^5$~GeV, $M_P \approx 10^{9.6}$~GeV and $M_U \approx 10^{16}$~GeV.}
\label{fig:coupl-unifn_PS}
\end{figure}

The general renormalization group equations (RGEs) can be found in Ref.~\cite{jones} and we here present the one-loop RGEs for the various gauge couplings
\begin{equation}
	\frac{d\alpha^{-1}_{i}}{dt} = -\frac{a_i}{2 \pi},
\label{rge-coupl}
\end{equation}
where we have denoted the fine structure constants by $\alpha_{i}=g^2_{i}/(4 \pi)$ with $i = 2L, 2R, 4C, 1R, BL, 3C, Y$, the one-loop beta coefficients by $a_i$ corresponding to the ${i^{\rm th}}$ gauge group and the log-scale of the energy by $t = \ln(\mu/\mu_0)$ with respect to an arbitrary reference energy $\mu_0$. The numerical values of $a_i$ for the breaking scheme (\ref{eq:BreakingChain}) is given in Ref.~\cite{PS-dparity}.

We find that gauge coupling unification is not consistent with TeV scale $W_R$, $Z_R$ gauge bosons and a reasonably low $D$ parity breaking scale around $10^{8}-10^{9}$ GeV. Few attempts have already been taken where $D$ parity is broken at higher scale ($> 10^{12}-10^{15}$ GeV) while keeping $W_R$ and $Z_R$ masses around few TeV scale \cite{0nubb-spatra-jhep, PS-dparity}. Since our subsequent analysis will based on the type II seesaw dominance mechanism for light neutrino masses, such large value ($> 10^{12}-10^{15}$ GeV) for $D$ parity breaking scale makes the type II seesaw contribution of neutrino mass really suppressed. 

Therefore, in addition to the usual Higgs multiplets $10_{H}$, $126_{H}$, $210_{H}$ and $54_{H}$, necessary for the breaking of $SO(10)$ to the SM, we have added extra $SO(10)$ Higgs representations, such as $16_H$ and $126_{H^\prime}$ to bring further down the scale of $D$ parity while maintaining the gauge coupling unification. The lowering of $D$ parity also increases the type-II seesaw contribution of neutrino mass as the latter is inversely proportional to the scale of $D$ parity breaking~\cite{Dparity-1}.

The symmetry breaking scales consistent with gauge coupling unification, and TeV scale RH gauge bosons $W_R$ and $Z_R$ are found to be $M_{B-L} \approx 4-5$~TeV, $M_{\Omega} \approx 6-10$~TeV, $M_{C} \approx 10^{5}-10^{6}$~GeV, $M_P \approx 10^{9.6}$~GeV and $M_U \approx 10^{16}$~GeV. The most desirable prediction of the model is the values of $g_{L}$ and $g_{R}$ at the TeV scale consistent with gauge coupling unification as $g_L \approx 0.63$ and $g_R \approx 0.38$. The ratio between these two couplings at the TeV scale is therefore
\begin{align}
	\frac{g_R}{g_L} = 0.60.
\end{align}
In the following sections, we will examine carefully the effect of $g_R = 0.6 g_L$ on the low energy process like $0\nu\beta\beta$ decay and LFV processes as well as collider signatures. Our model is not the only possibility to produce a  non-universality between the left and right gauge couplings at the TeV scale. A survey of models based on SO(10) without manifest Left-Right symmetry is presented in \cite{romao_paper}.

\subsection{Neutrino Masses}
The relevant terms responsible for giving masses to the three generations of leptons are
\begin{align}
	\mathcal{L}_{yuk} = h_{ij}\overline{\ell_{Li}}\ell_{Rj}\Phi+
	\tilde{h}_{ij}\overline{\ell_{Li}}\ell_{Rj}\tilde{\Phi} 	
	+ f_{ij}\left[\overline{(\ell_{Li})^c}\ell_{Lj}\Delta_L+
	\overline{(\ell_{Ri})^c}\ell_{Rj}\Delta_R\right] + \text{h.c.},
\label{yukaw}
\end{align}
where $\ell_{L,R}^T = (\nu_{L,R}, e_{L,R})$. The discrete left-right symmetry ensures that the Majorana Yukawa coupling matrix $f$ is the same for both left- and right-handed neutrinos. The breaking of left-right symmetry to $U(1)_{em}$ results in the complete $6 \times 6$ neutral lepton mass matrix $M_\nu$ in the $(\nu_L, N^c_R)$ basis given as 
\begin{equation}
	M_\nu= 
	\left(\begin{array}{cc}
		M_L  & M_D   \\
   	M^T_D & M_R
\end{array} \right),
\label{eqn:numatrix}       
\end{equation}
where the Majorana mass matrix $M_R$ of the heavy RH neutrinos is dynamically generated by the VEV of RH heavy Higgs triplet, i.e $\langle \Delta_R \rangle = v_R$ while the Majorana mass matrix of the LH neutrino masses is generated through the Higgs triplet VEV $\langle \Delta_L \rangle = v_L$, $M_{R,L} = f v_{R,L}$. The Dirac mass matrix is given by $M_D = h v_u + \tilde{h} v_d$. The symmetric complex matrix $M_\nu$ is diagonalized by a unitary mixing matrix $\mathcal{V}$ relating the neutrino flavour states with the mass eigenstates ($\alpha, \beta = e, \mu, \tau$ denote the flavour states while $i,j,k = 1, 2, 3$ denote the mass eigenstates) as
\begin{align}
	\begin{pmatrix}
		\nu_{\alpha L} \\
		N^c_{\beta R}
	\end{pmatrix} = 
	\mathcal{V} 
	\begin{pmatrix}
		\nu_i \\
		N_k
	\end{pmatrix}=
	\begin{pmatrix}
		U & S \\
		T & V
	\end{pmatrix}
	\begin{pmatrix}
		\nu_i\\
		N_k
	\end{pmatrix}.
\end{align}
The diagonalization is expressed as $M_\nu^\text{diag} \equiv \text{diag}(m_1, m_2, m_3, M_1, M_2, M_3) = \mathcal{V}^\dagger M_\nu \mathcal{V}^*$ and it can be understood in two steps: firstly, the block diagonalization of $M_\nu$ by a mixing matrix $\mathcal{W}$ as $\mathcal{W}^\dagger M_\nu \mathcal{W}^* = M_\text{BD}$, and secondly, the block diagonalized mass matrix is diagonalized by a mixing matrix $\mathcal{U}$ as $\mathcal{U}^\dagger M_\text{BD} \mathcal{U}^* = M_\nu^\text{diag}$. The form of the unitary mixing matrix $\mathcal{V}$ is to first order in the left-right mixing given by
\begin{equation}
\label{eq:Wmatrix}
	\mathcal{V} \equiv \mathcal{W} \cdot \mathcal{U} \equiv 
	\begin{pmatrix} 
		U & S \\ 
		T & V 
	\end{pmatrix} 
	\simeq 
	\begin{pmatrix} 
		\mbox{1} - \frac{1}{2}RR^\dagger & R \\ 
		-R^\dagger & \mbox{1} - \frac{1}{2}R^\dagger R
	\end{pmatrix}
	\cdot
	\begin{pmatrix} 
		U_\nu & 0 \\ 
		0     & U_N 
	\end{pmatrix}
	\approx
	\begin{pmatrix} 
		U_\nu & R\, V \\
		-R^\dagger U_\nu & V
	\end{pmatrix} + \mathcal{O}(R^2),
\end{equation}
with $R = M_D M_R^{-1} + {\cal O}(M_D^3(M_R^{-1})^3)$ defining the left-right mixing. The last approximation is valid for the case of small left-right mixing i,e $R = M_D M^{-1}_R \ll 1$. We now have $\mathcal{V}^\dagger M_\nu \mathcal{V}^* = M_\nu^\text{diag}$, where the unitary matrices $U_\nu$ and $U_N$ are defined by 
\begin{align}
	M_L - M_D M_R^{-1} M_D^T &= U_\nu \cdot \text{diag}(m_1,m_2,m_3) \cdot U_\nu^T, \\
	M_R &= U_N \cdot \text{diag}(M_1, M_2, M_3) \cdot U_N^T.
\label{eq:mnu_MR_def}
\end{align}
In the basis where the charged lepton mass matrix is already diagonal, the matrix $U_\nu$ can be identified with left-handed charged-current mixing matrix $U_{\rm PMNS}$. With these simplifications, the light neutrino mass matrix can be written as
\begin{align}
	m_\nu = M_L - R M_R R^T
	      = m_\nu^{II} + m_\nu^I,
\label{neutrino-mass}
\end{align}
where $m_\nu^{II} = M_L = f v_L$ is the type-II seesaw contribution \cite{type-II-group} and $m_\nu^{I} = - M_D M_R^{-1} M_D^T = - R M_R R^T$ is the type-I seesaw contribution \cite{type-I-group} to the light neutrino masses. Observed phenomenology requires $v_L \ll v_d < v_u \ll v_R$. The analytic expression for the VEV of the LH Higgs triplet $\Delta_L$, by minimizing the scalar potential, can be expressed as~\cite{Dparity-1}
\begin{equation}
	v_L \approx \frac{\beta v^2v_R}{M\langle \sigma \rangle},
\label{vev_value} 
\end{equation}
where $v=\sqrt{v_u^2+v_d^2} \approx 246$~GeV, $\beta$ is a coupling constant of $\mathcal{O}(1)$ and the other mass scales $M, \langle \sigma \rangle$ are of the order of the $D$ parity breaking scale i.e. $M_P \simeq 10^{8}-10^{9}$~GeV for our present discussion. Notice that in the above equation the smallness of the VEV of $\Delta_L$ is determined by the $D$ parity breaking scale, not the $U(1)_R \times U(1)_{B-L}$ breaking scale and hence there are no constraints on $v_R$ from the type-II seesaw point of view. As a result the mass scale of RH neutrinos can be of the order of the TeV scale while consistent with the oscillation data. 

It is worth mentioning here that the matrix $R$ plays a crucial role for phenomenology as it describes the mixing between the light LH and the heavy RH neutrinos. However, we here restrict our analysis to the case where the light neutrino mass matrix is governed by pure type-II seesaw dominance while assuming the Dirac neutrino mass matrix is small and therefore the type-I contribution plays a negligible role. See for instance~\cite{0nubb-tello-2011}. Under the assumption of type-II seesaw dominance, the light neutrino mass matrix is directly proportional to the heavy neutrino mass matrix,
\begin{eqnarray}
	m_\nu \approx M_L = f v_L = \frac{v_L}{v_R} M_R = 
	\frac{\beta v^2}{M \langle \sigma \rangle} M_R.
\end{eqnarray}
An immediate important consequence is that the mass spectra are directly proportional to each other, i.e. $m_\nu \propto M_R$.

\subsection{Charged-Current Lepton Interactions}
Most relevant for our phenomenological analysis, the charged current interactions in the flavour basis valid for this particular version of a left-right model at the TeV scale with unequal strength of $SU(2)_L$ and $SU(2)_R$ gauge couplings are
\begin{align}
	{\cal L}^{\rm lep}_{CC} &= \frac{g_L}{2\sqrt{2}} \sum_{\alpha=e, \mu, \tau}
	\left[
	\overline{\ell_{\alpha_L}} \gamma^\mu (1-\gamma_5) 
	\nu_{\alpha_L} W_{L_\mu}
	+ \frac{g_R}{g_L} \overline{\ell_{\alpha_R}} \gamma^\mu (1+\gamma_5) 
	N_{\alpha_R} W_{R_\mu}
	\right] +\text{h.c.} \nonumber \\
	&\approx \frac{g_{L}}{\sqrt{2}} \sum_{i=1}^3 
	\left[
	U_{ei} \overline{e_L}\gamma^{\mu} \nu_{Li}(W_{1\mu}^- + \xi W_{2\mu}^-) 
  + \frac{g_R}{g_L}	
  V^*_{ei} \overline{e_R}\gamma^{\mu} N_{Ri} (-\xi W_{1\mu}^- + W_{2\mu}^-)
	\right]
  + \text{h.c.},
\label{eq:form-CC-flavor}
\end{align}
where we have considered that the LH and RH charged gauge bosons mix with each other and hence the physical gauge bosons are linear combinations of $W_L$ and $W_R$ as
\begin{align}
\label{eqn:LRmix} 
		W_1 &= \phantom{-} W_L \cos\xi  + W_R \sin\xi,  \\ 
		W_2 &= -           W_L \sin\xi  + W_R \cos\xi 
\end{align}
with
\begin{equation}
	|\tan 2\xi| \sim \frac{v_u v_d}{v^2_R} \sim \frac{v_d}{v_u} 	
	\left(\frac{g_R}{g_L}\right)^2 
	\left (\frac{M_{W_L}}{M_{W_R}}\right)^2 \,,
\end{equation}
where the upper bound on the mixing angle is given by $\xi \lsim \left (\frac{M_{W_L}}{M_{W_R}}\right)^2$. For a TeV scale right handed gauge boson one gets $\xi \lsim 10^{-3}$~\cite{umasankar, zarlek}. As mentioned above, here we consider a case where the Dirac term in the neutrino mass matrix is negligible in order to have type-II seesaw dominance. Since the type-II seesaw mechanism gives a light neutrino mass matrix of the form $m_\nu =f v_L = \frac{v_L}{v_R} M_R$, the mixing among the LH and RH fields are equal up to a phase, i.e $V = U^*$.

\section{Neutrinoless Double Beta Decay}
\label{sec:0nu2beta}

The purpose of this section is to examine analytically as well as numerically in detail the relevant contributions to $0\nu \beta \beta$ decay within the model under consideration. 

\subsection{Analytic Amplitudes}

\begin{figure}[t]
\centering
\includegraphics[width=0.30\columnwidth]{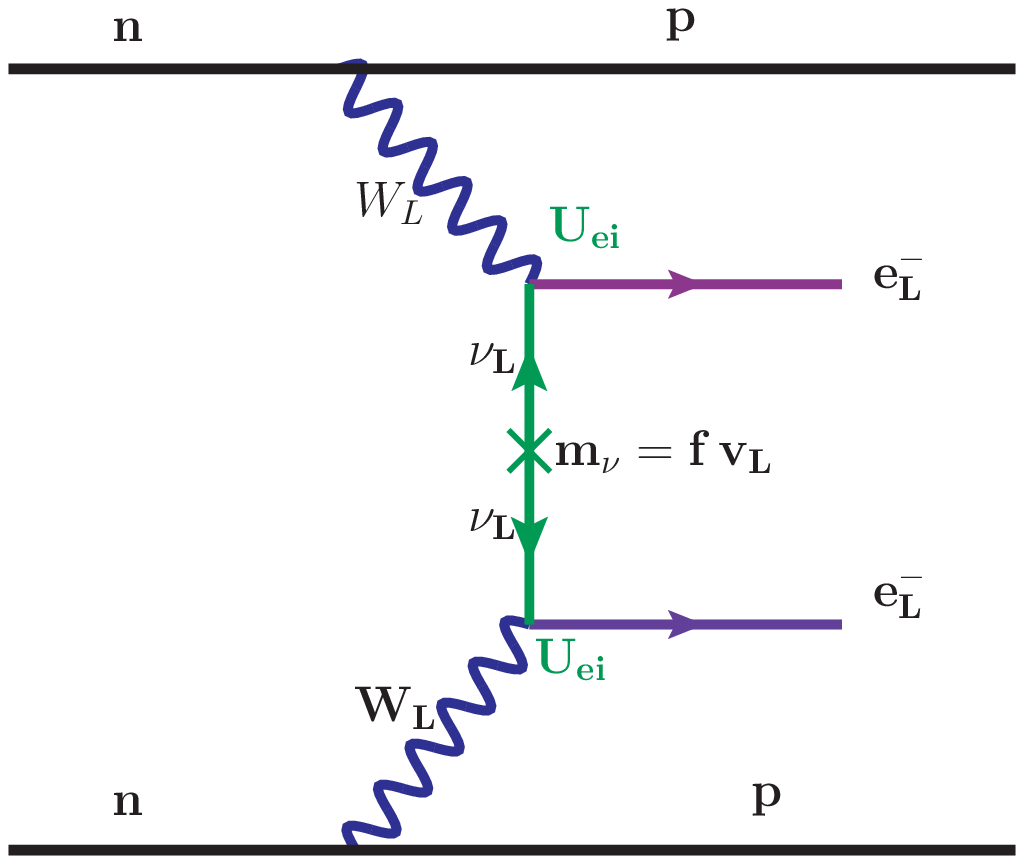}
\includegraphics[width=0.30\columnwidth]{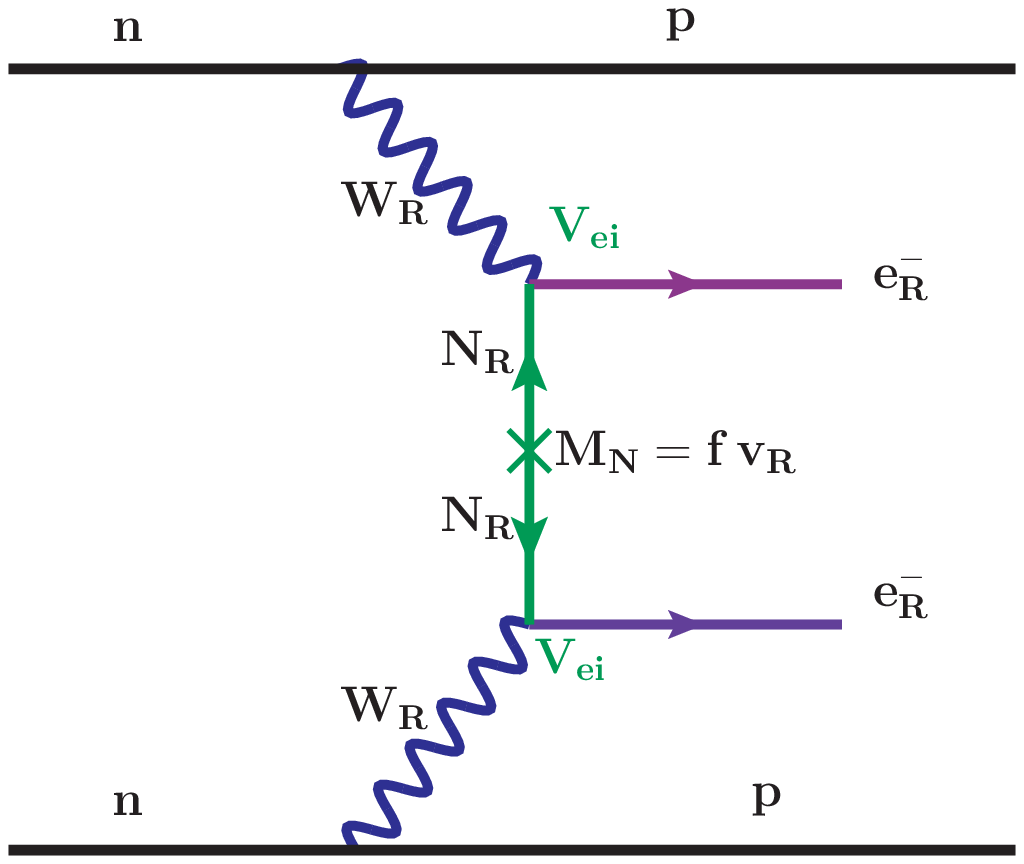}
\includegraphics[width=0.34\columnwidth]{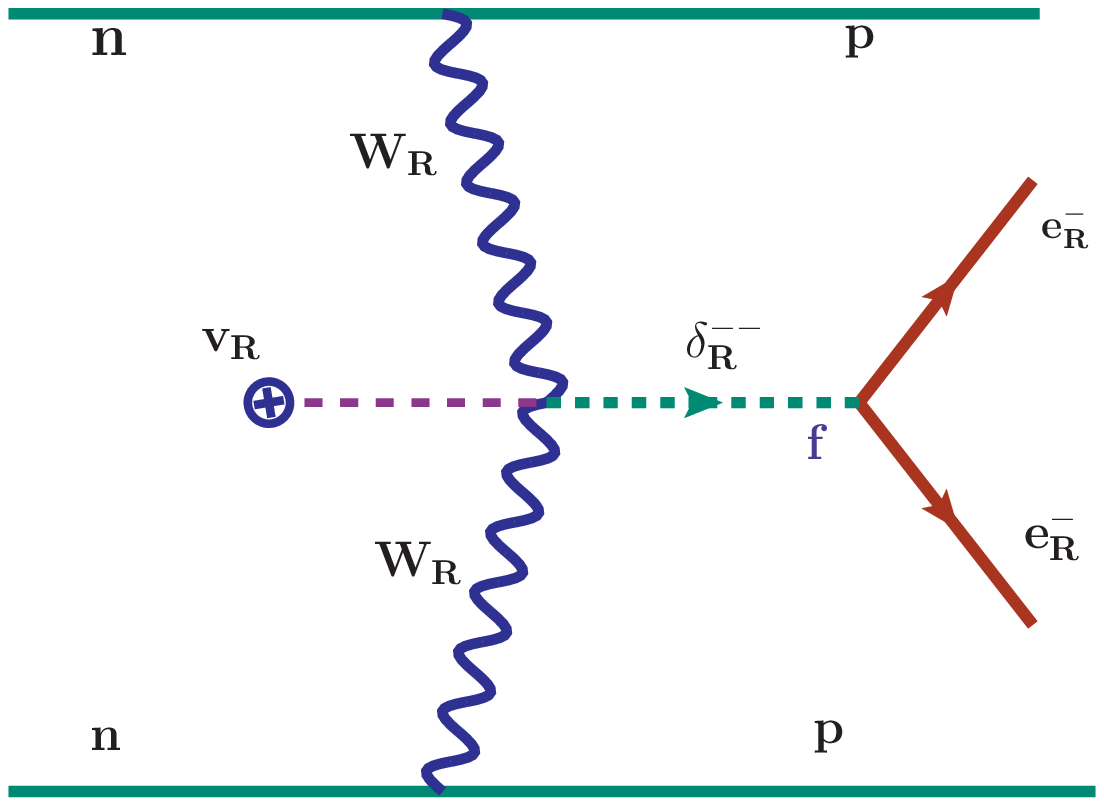}
\caption{Diagrams contributing to neutrinoless double beta decay: Standard LH charged current interaction through the exchange of light Majorana neutrinos (left), RH charged current interaction through the exchange of heavy $W_R$ bosons and heavy neutrinos (middle), doubly charged RH triplet Higgs scalar mediation (right).}
\label{fig:Feyn-type-II-nu}
\end{figure}
Following the form of the charged current interactions given in Eq.~\eqref{eq:form-CC-flavor}, there are several Feynman diagrams contributing to $0\nu\beta\beta$ process. These diagrams include (i) the standard mechanism due to $W_L - W_L$ mediation via the exchange of light active Majorana neutrinos $\nu_L$ (Fig.~\ref{fig:Feyn-type-II-nu}~(left)), (ii) $W_R - W_R$ mediation with purely RH charged current interaction through the exchange of heavy RH Majorana neutrinos (Fig.~\ref{fig:Feyn-type-II-nu}~(middle)), (iii) exchange of charged triplet Higgs fields. The diagram with $W_R - W_R$ mediation via the RH Higgs triplet exchange is shown in Fig.~\ref{fig:Feyn-type-II-nu}~(right) while the analogous diagram with a LH Higgs triplet is severely suppressed by the light neutrino masses; (iv) diagrams arising due to the affect of $W_L - W_R$ mixing which are suppressed in our case as the $W_L-W_R$ mixing is found to be $< 10^{-3}$, (iv) diagrams arising due to light-heavy neutrino mixing proportional to $\frac{M_D}{M_R}$ and therefore negligible in our case. The Feynman amplitudes for the three contributions in Fig.~\ref{fig:Feyn-type-II-nu} can be expressed as
\begin{align}
\label{eq:0nubb_ampl_mm}
	\mathcal{A}_{\nu} &\approx 
		  \phantom{-} G^2_F \sum_{i=1}^3 \frac{U^2_{e i}  m_i}{p^2} 
			\equiv G^2_F  \frac{m_{\rm ee}}{p^2}, \\
	\mathcal{A}_N &\approx 
		- G^2_F \left(\frac{g_R}{g_L}\right)^4 \left(\frac{M_{W_L}}{M_{W_R}}\right)^4 
		\sum_{i=1}^3 \frac{V^2_{e i}}{M_{N_i}}, \\
	\mathcal{A}_{\delta^{--}_R} &\approx 
		- G^2_F \left(\frac{g_R}{g_L}\right)^4 \left(\frac{M_{W_L}}{M_{W_R}}\right)^4
		\sum_{i=1}^3 \frac{V^2_{e\,i}\, M_{N_i}}{M^2_{\delta^{--}_R}}.
\end{align}
where $|p|\simeq 100$~MeV is a measure of the light neutrino momentum transfer, of the order of the nuclear scale and $G_F$ is the Fermi coupling constant. Equation~\eqref{eq:0nubb_ampl_mm} defines the effective $0\nu\beta\beta$ mass $m_\text{ee}$. The analytic expression for the Feynman amplitude due to heavy neutrino exchange has been derived assuming that the heavy neutrino masses are larger than the typical momentum exchange scale, i.e. $M_i \gg |p|$.

For a rough estimate of the individual contributions, we define the masses of all particles belonging to the RH sector ($M_{N_i}, M_{W_R}, M_{\delta^{--}_R}$) as $\Lambda_R \approx$~TeV (although $M_{\delta^{--}_R}$ can be as light as a few hundred GeV) and the mass of $W_L$is of order $\Lambda_\text{EW} \approx 100$~GeV. With these approximations and using $V = U$and $|p|=100$~MeV, one can compare the non-standard amplitudes due to heavy RH Majorana neutrino and doubly charged Higgs exchange with the standard mechanism as
\begin{align}
\label{eq:comp:amplNR}
	\left|\frac{\mathcal{A}_{N_R}}{\mathcal{A}_{\nu}}\right| \approx
	\left|\frac{\mathcal{A}_{\delta^{--}_R}}{\mathcal{A}_{\nu}}\right| \approx 
	\left(\frac{g_R}{g_L}\right)^4
	\frac{|p^2|}{m_\nu} \frac{\Lambda_\text{EW}^4}{\Lambda_R^5}.
\end{align}
It is clear that the non-standard contributions can dominate over the standard mechanism for TeV scale RH particles.

\subsection{Decay Rates}

%
\begin{table}[t]
\centering
\vspace{10pt}
\begin{tabular}{rccc}
\hline
Isotope & $G^{0\nu}_{01}$ (y$^{-1}$) & ${\cal M}_\nu$ & ${\cal M}_N$  \\
\hline 
$^{76}$Ge   & $5.77 \times 10^{-15}$ & 2.58--6.64 & 233--412  \\ 
$^{136}$Xe  & $3.56 \times 10^{-14}$ & 1.57--3.85 & 164--172  \\
\hline
\end{tabular}
\caption{Phase space factor $G^{0\nu}_{01}$~\cite{phase-space-0nubb} and nuclear matrix elements ${\cal M}_\nu$, ${\cal M}_N$~\cite{phase-NMEs-0nubb} for the isotopes $^{76}$Ge and $^{136}$Xe. The NMEs of $^{76}$Ge have been used throughout our numerical calculation, including their indicated uncertainties as appropriate.}
\label{tab:nucl-matrix}
\end{table}
The Feynman diagrams for $0\nu\beta\beta$ transition as shown in Fig.\ref{fig:Feyn-type-II-nu} have LH ($e_L^- e^-_L$) or RH ($e_R^- e^-_R$) electrons in the final states. Thus, the corresponding nuclear matrix elements affect the decay rates of different particle exchange contributions to $0\nu\beta\beta$ decay which crucially depends upon the chirality of the hadronic and leptonic currents involved \cite{0nubb-vergados, 0nubb-rod-barry, 0nubb-psb-lee, 0nubb-spatra-jhep}. The analytic expression for the inverse half-life of a given nuclear isotope considering only the relevant contributions due to light neutrino, heavy neutrino and doubly charged RH Higgs can be expressed as 

\begin{align}
\frac{1}{T_{1/2}^{0\nu}} =
	G^{0\nu}_{01} 
	\left| 
		{\cal M}_\nu \eta_\nu + {\cal M}_N(\eta_{N_R}+\eta_{\delta_R})
	\right|^2 
 	\approx G^{0\nu}_{01}
	\left(
		|{\cal M}_\nu|^2|\eta_\nu|^2 + |{\cal M}_N|^2 
		\left( |\eta_{N_R}|^2  + |\eta_{\delta_R}|^2 \right)
  \right),
\label{eq:half-life_typeII}
\end{align}
where $G^{0\nu}_{01}$ is the nuclear phase space factor in the standard mechanism and ${\cal M}_{\nu,N}$ are the respective nuclear matrix elements (NMEs). The numerical values for these quantities are presented in Table.~\ref{tab:nucl-matrix} for isotopes $^{76}$Ge and $^{136}$Xe. In Eqn.~\eqref{eq:half-life_typeII}, we neglect small interference terms between the different contributions. The dimensionless particle physics parameters $\eta_X$ characterizing are given by
\begin{align}
\label{eqn:eta1}
	\eta_{\nu} &= \frac{1}{m_e}\sum_{i} U_{ei}\, m_i, \\
\label{eqn:eta2}
	\eta_{N_R} &= m_p \left(\frac{g_R}{g_L}\right)^4  \left(\frac{M_{W_L}}{M_{W_R}}\right)^4\, 
     \sum_{i} \frac{V^{*2}_{ei}}{M_{N_{i}}}, \\
\label{eqn:eta3}
	\eta_{\delta_R} &= m_p \left(\frac{g_R}{g_L}\right)^4  
	\left(\frac{M_{W_L}}{M_{W_R}}\right)^4\, 
     \sum_{i} \frac{V^{*2}_{ei} M_{N_i}}{M_{\delta^{--}_R}^2},
\end{align}
where $m_e$ and $m_p$ is the mass of electron and proton, respectively. 

The experimental non-observation of $0\nu\beta\beta$ currently provides lower bounds on the half life of this process in various isotopes. So far there has been only one claim of observation of $0\nu\beta\beta$ with $T_{1/2} \approx 2.23^{+0.44}_{-0.31} \times 10^{25}$~yr in ${}^{76}$Ge at 68\% CL by a part of Heidelberg-Moscow (HM) experiment~\cite{Klapdor-claim}. Using the same isotope, the non-observation of $0\nu\beta\beta$ by GERDA collaboration recently sets a new limit on the half-life to be $T_{1/2} ({}^{76} {\rm Ge}) > 2.1 \times 10^{25}$~yr at 90\% CL \cite{gerda}. The combined limit from all the Ge based experiments such as HM~\cite{Klapdor-bound}, GERDA~\cite{gerda} and IGEX \cite{Igex} gives $T_{1/2}({}^{76} {\rm Ge}) > 3.0 \times 10^{25}$~yr at 90\% CL, which strongly disfavours earlier claim of HM experiment~\cite{Klapdor-claim}.

Similarly the non-observation of $0\nu\beta\beta$ by the KamLAND-Zen Collaboration recently provide a lower bound on the half life of ${}^{136}$Xe to be $T_{1/2} > 1.9 \times 10^{25}$~yr \cite{kamland}. In case of ${}^{76} {\rm Ge}$, the combined limit on half-life time from HM~\cite{Klapdor-bound}, GERDA~\cite{gerda} and IGEX \cite{Igex} can be translated into bounds for the effective parameters Eqs.~\eqref{eqn:eta1} - \eqref{eqn:eta3},
\begin{align}
	\eta_\nu            \lesssim (3.6 - 9.3) \times 10^{-7}, \quad
	\eta_{N_R,\delta_R} \lesssim (0.6 - 1.0) \times 10^{-8}, 
\end{align}
assuming only one contribution is dominant at a time.

\begin{figure}[t]
\centering
\includegraphics[scale=.34,angle=-90]{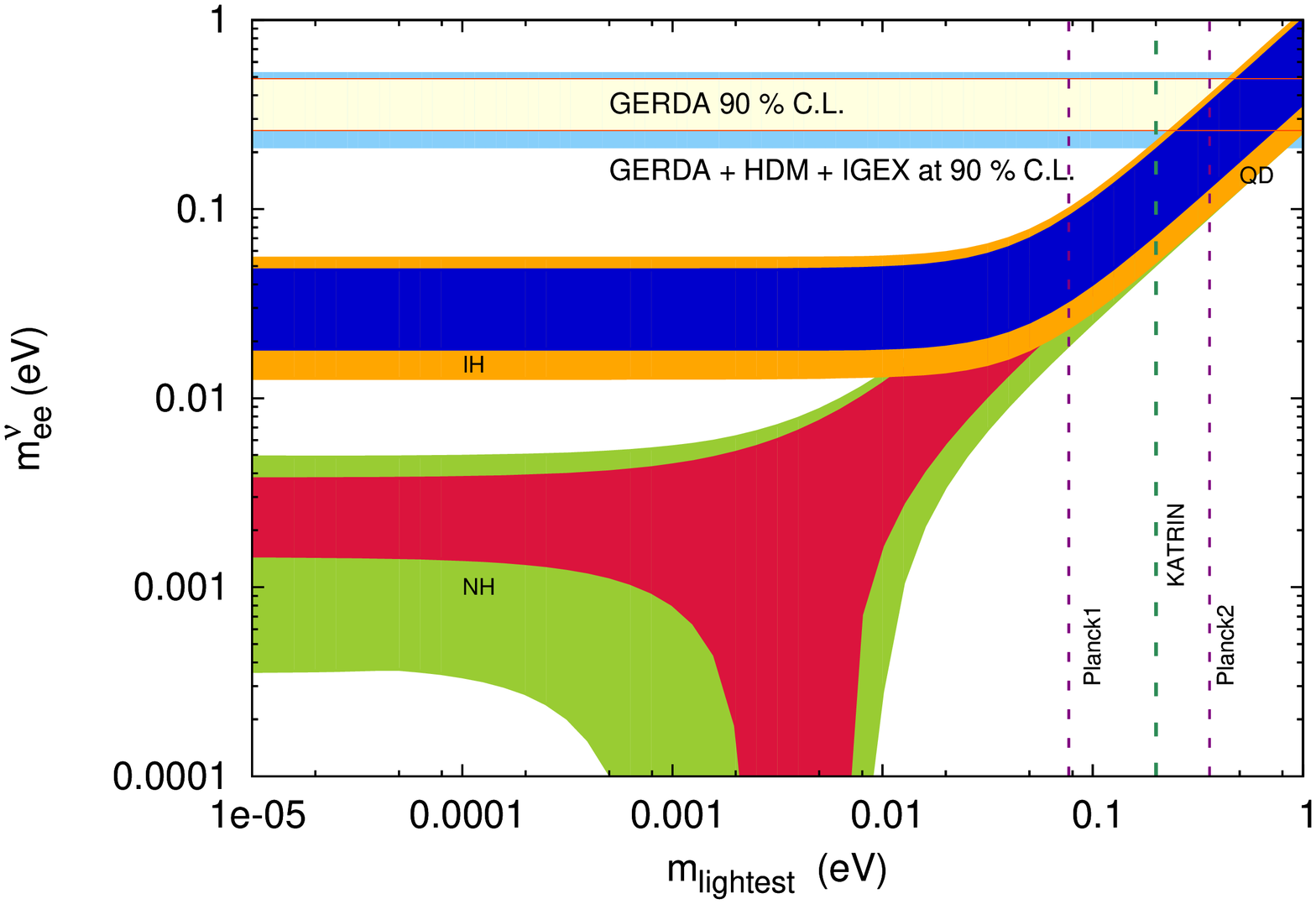}
\includegraphics[scale=.34,angle=-90]{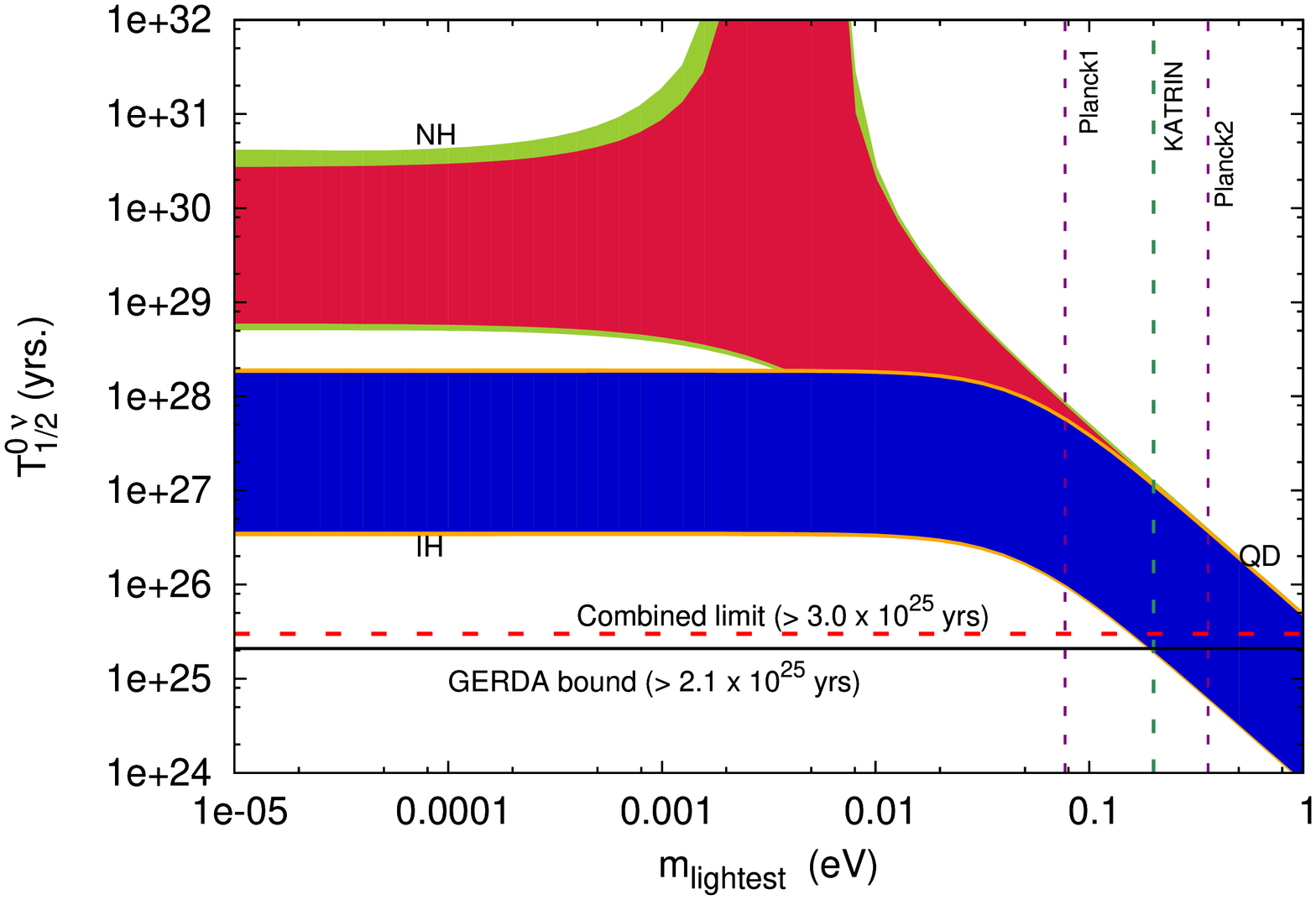}
\caption{Effective $0\nu\beta\beta$ mass (left) and $^{76}$Ge half life (right) in the standard mechanism as a function of the lightest neutrino mass. The red (blue) coloured bands show the variation due the Majorana phases for a NH (IH) pattern of light neutrinos using the best-fit values of the oscillation parameters while the green and yellow bands correspond to an additional $3\sigma$ variation of the oscillation parameters.The vertical lines represent bounds from Planck and KATRIN while the horizontal lines denote the experimental bounds from GERDA and a combination of the GERDA, Heidelberg-Moscow and IGEX experiments.}
\label{typeII-lr-std-nu}
\end{figure}
The analytic expression of the effective mass parameter due to the exchange of light neutrinos can be written in terms of the PMNS mixing angles $\theta_{12}$, $\theta_{13}$, the Majorana phases $\alpha$, $\beta$ and the mass eigenvalues with a NH ($m_1 < m_2 \ll m_3$) or an IH pattern ($m_3 < m_1 \ll m_2$),
\begin{align}
	m^\nu_\text{ee} = \left|
		  m_1 c_{12}^2 c_{13}^2 
		+ m_2 s_{12}^2 c_{13}^2 e^{-i \alpha} 
		+ m_3\, s_{13}^2 e^{-i \beta} 
	\right|.
\label{eq:mee-std-nu} 
\end{align}
Fig.~\ref{typeII-lr-std-nu} shows the prediction for the effective $0\nu\beta\beta$ mass $m^\nu_{ee}$ (left panel) and the corresponding half life of $^{76}$Ge (right panel) as a function of the lightest neutrino mass. The coloured bands show the variation due to the experimental uncertainty in the oscillation parameters and the unknown Majorana phases $0 \leq \alpha, \beta \leq \pi$. The bands in Fig~\ref{typeII-lr-std-nu}~(right) in addition include the $^{76}$Ge NME uncertainties as shown in Tab.~\ref{tab:nucl-matrix}. The red (blue) coloured bands correspond to a NH (IH) pattern of light neutrinos using the best-fit values of the oscillation parameters while the green and yellow bands correspond to a $3\sigma$ variation of the oscillation parameters, as given in Tab.~\ref{table-osc}. The limit on the half life of $^{76}$Ge by the GERDA experiment can be translated to a bound on the effective mass parameter $m^\nu_{ee} \leq 0.21 - 0.53$ eV, presented by the dashed blue lines shaded by a light blue band in the left panel of Fig.~\ref{typeII-lr-std-nu}. Similarly, the corresponding bound on the effective mass can be derived from the combined experimental limits from GERDA, Heidelberg-Moscow and IGEX at 90\% C.L and is denoted here by the corresponding horizontal line and band. The vertical dashed lines denote the experimental bounds on the lightest neutrino mass from cosmology \cite{planck} and KATRIN \cite{Katrin}. 
\begin{table}[t!]
\begin{tabular}{ccc}
\hline
Oscillation parameter & Best Fit & 3$\sigma$ Range \\
\hline
$\Delta m^2_{21}  [10^{-5} \text{eV}^2]$              & 7.500 & 7.00 - 8.09 \\
$|\Delta m^2_{31} (\text{NH})| [10^{-3} \text{eV}^2]$ & 2.473 & 2.27 - 2.69 \\
$|\Delta m^2_{23} (\text{IH})| [10^{-3} \text{eV}^2]$ & 2.420 & 2.24 - 2.65 \\
$\sin^2\theta_{12}$                                   & 0.306 & 0.27 - 0.34 \\
$\sin^2\theta_{23}$                                   & 0.420 & 0.34 - 0.67 \\
$\sin^2\theta_{13}$                                   & 0.021 & 0.016- 0.03 \\
\hline
\end{tabular}
\caption{Global best fit values and 3$\sigma$ uncertainties for the neutrino mass squared differences and mixing angles \cite{schwetz12}.}
\label{table-osc}
\end{table}  

From Fig.~\ref{typeII-lr-std-nu} it can be inferred that if one assumes that only the light Majorana neutrinos contribute to $0\nu\beta\beta$ and the light neutrinos are quasi-degenerate, then the experimental limit on the half-lives \cite{gerda, kamland} is saturated as shown in the right panel of Fig.~\ref{typeII-lr-std-nu}. However, the stringent bound on the sum of masses of the light neutrinos, i.e $\sum_i m_i \leq 0.23$~eV (95\% C.L.) from astrophysical observation such as Planck~\cite{planck}, is in tension with a QD nature of light neutrinos. Future sensitivities of the planned $0\nu\beta\beta$ decay experiments aimed at probing the effective mass less than $0.1$ eV, may probe IH pattern of light neutrinos. In order to probe the normal hierarchy is not possible in the near future with $0\nu\beta\beta$ decay experiments and moreover, it is possible that the decay rate might effectively vanish because of the presence of the Majorana phases. While the tension between cosmology and $0\nu\beta\beta$ is not yet stringent, one possibility to evade it is through the presence of non-standard new physics contributions to $0\nu\beta\beta$ decay, for example arising in left-right symmetry models as discussed here.

For type-II seesaw dominance, the mass eigenvalues of the RH Majorana neutrinos are directly proportional to the light neutrino masses,
\begin{equation}
	M_{N_i} = \frac{v_R}{v_L} m_{\nu_i},
\label{eq:typeII_dom_eigen_rel:a}
\end{equation}
where the ratio of the two VEVs $\frac{v_R}{v_L}$ is independent of the generation. Hence, one can express the heavy neutrino masses in terms of light neutrino masses, for example fixing the heaviest RH neutrino mass at 1~TeV (i.e. $M_{N_3}$ for NH pattern and $M_{N_2}$ for IH pattern),
\begin{align}
	M_{N_i} &= \frac{m_i}{m_{3}} M_{N_3},\text{ NH}, 
\label{eq:NH_massrel-typeII} \\
	M_{N_i} &= \frac{m_i}{m_{2}} M_{N_2},\text{ IH}.
\label{eq:IH_massrel-typeII} 
\end{align}
Similarly, the heavy neutrino mixing matrix is determined by $V = U^*$ and consequently all three $0\nu\beta\beta$ effective parameters $\eta_\nu$, $\eta_{N_R}$ and $\eta_{\delta_R}$ can be expressed in terms of the light neutrino oscillation parameters, the Majorana phases $\alpha$, $\beta$, the lightest neutrino mass and the heavy masses $M_{W_R}$, $M_{\delta^{--}_R}$, $M_{N_{2,3}}$. In addition, the RH contributions to the $0\nu\beta\beta$ half life are suppressed by $(g_R/g_L)^8$ leading to reduction of almost two orders of magnitude compared to the minimal left-right symmetric model with $g_R = g_L$ and all other parameters kept the same. 

\begin{figure}[t]
\centering
\includegraphics[scale=0.38,angle=-90]{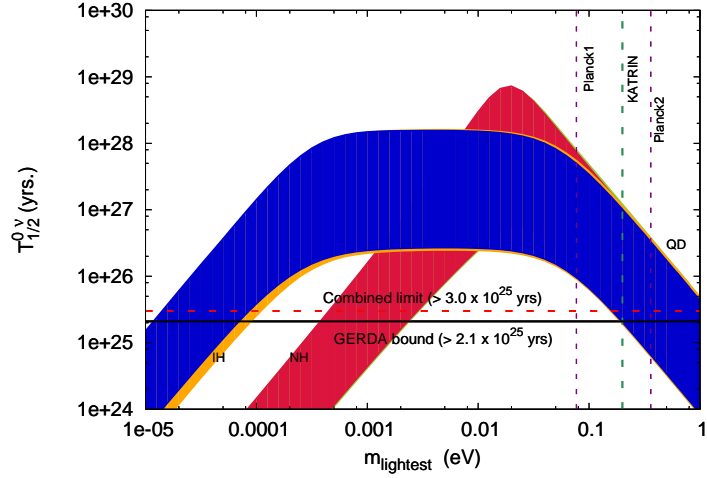}
\caption{As Fig.~\ref{typeII-lr-std-nu}, but showing the $0\nu\beta\beta$ half life of $^{76}$Ge arising in the combined contribution from light neutrino, heavy neutrino and RH Higgs triplet exchange. The heavy mass scales are fixed as $M_N = M_{\delta_{R}} = 0.75$~TeV and $M_{W_R}=2.2$~TeV.}
\label{typeII-half-life-combined}
\end{figure}
In Fig.~\ref{typeII-half-life-combined}, we show the $^{76}$Ge $0\nu\beta\beta$ half-life  due to the combination of light neutrino, heavy neutrino and RH Higgs triplet exchange as a function of the lightest neutrino mass. In order to estimate this half life we have taken the best fit and $3\sigma$ ranges of oscillation parameters from Table.~\ref{table-osc}, the NMEs from Tab.~\ref{tab:nucl-matrix} and fixed the heavy mass scales as $M_N = M_{\delta_{R}} = 0.75$~TeV, $M_{W_R}=2.2$~TeV. In addition to the oscillation parameters, the Majorana phases have been varied from $0$ to $\pi$. Several important conclusions can be drawn: (i) the purely RH current contributions via heavy neutrino and triplet Higgs exchange can saturate the current experimental limit even for a normal hierarchical pattern of light neutrinos. (ii) one can derive a lower bound on the lightest neutrino mass of $m_\text{lightest} \approx$ 0.45 - 3.5~meV (NH pattern) and 0.015 - 0.1~meV (IH pattern) due to the heavy RH contributions increasing for smaller $m_\text{lightest}$. This limit depends on the heavy mass scales. (iii) There is an upper limit on the half life, i.e. it is not possible that all contributions vanish at the same time.
       
We conclude that the current experimental bounds and near future sensitivity in lepton number violating $0\nu\beta\beta$ decay can also be saturated in a normal hierarchical pattern of light neutrinos due to the non-standard contributions from heavy RH neutrino and RH Higgs triplet exchange. Hence, if a future experiment observes a signal of $0\nu\beta\beta$ decay with a half-life of $T^{0\nu\beta\beta}_{1/2} \approx 3 \times 10^{26}$~y, it could be an indication for beyond the SM physics.

\begin{figure}[t!]
\centering
\includegraphics[width=0.34\textwidth,angle=-90]{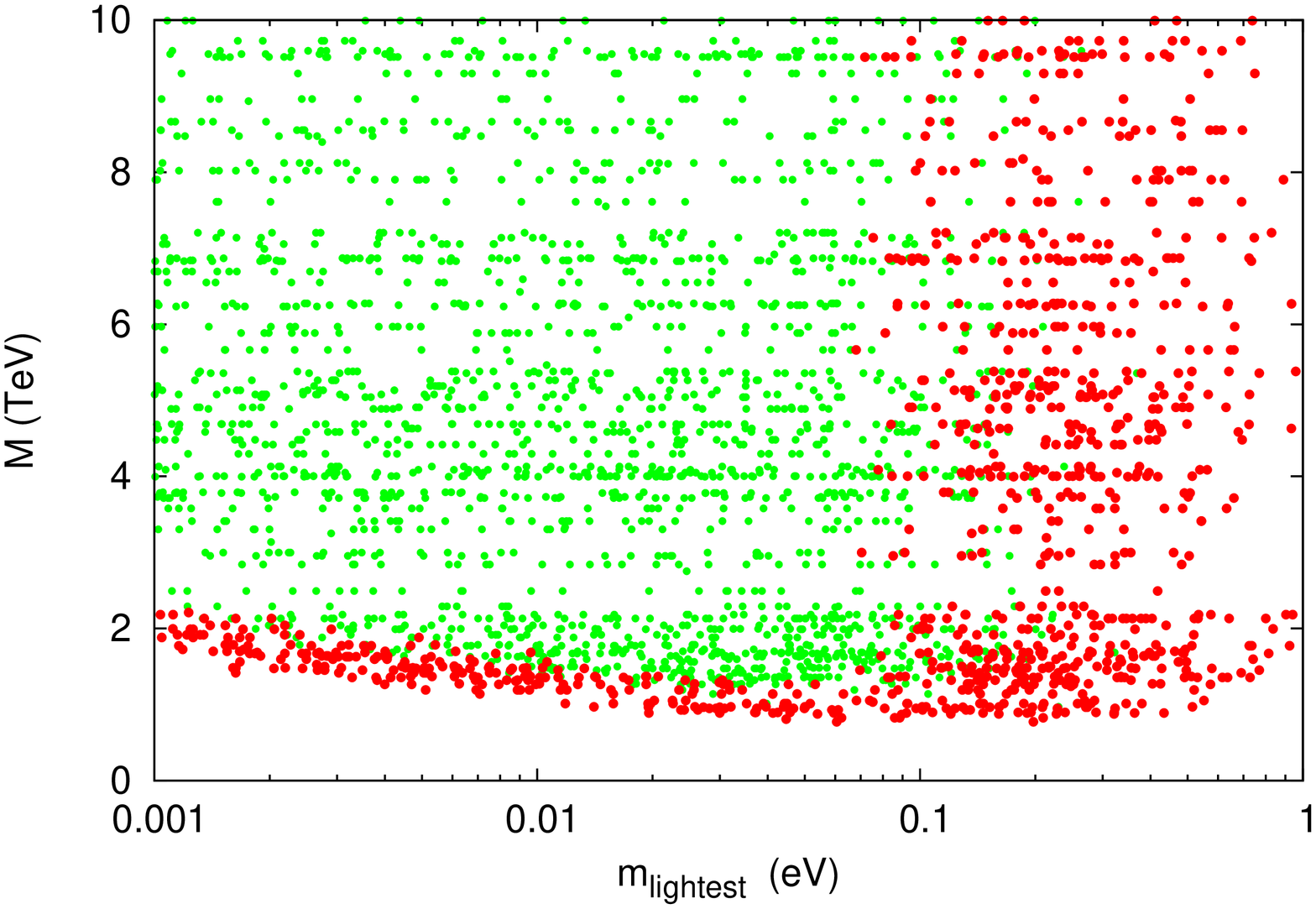}
\includegraphics[width=0.34\textwidth,angle=-90]{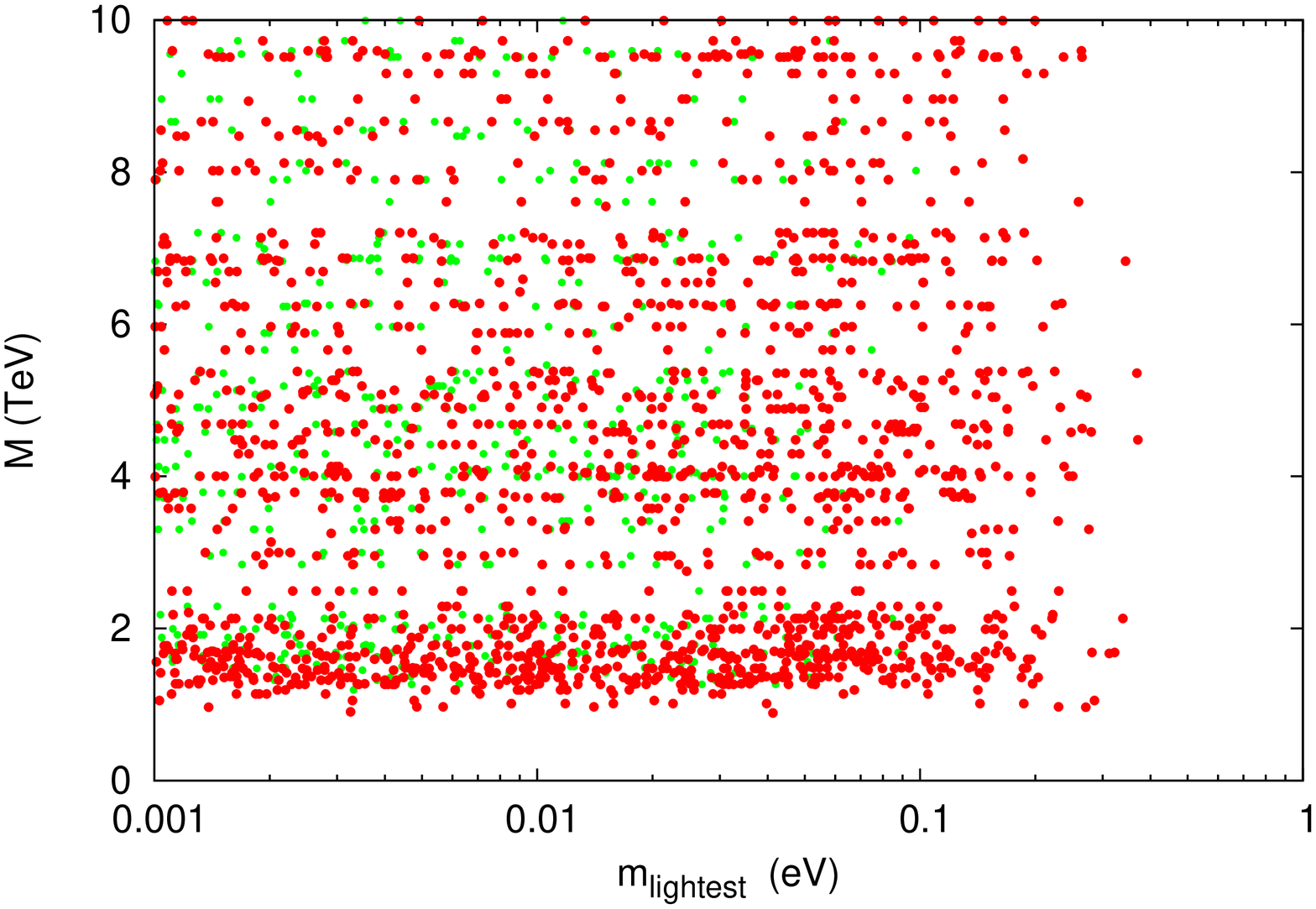}
\caption{Scatter plot in the common heavy scale $M \equiv M_{W_R} \simeq M_\Delta \simeq M_N$ and the lightest neutrino mass ($m_{\rm lightest}$), for NH (left) and IH (right) light neutrinos. The light green dots represent parameter points satisfying the current $0\nu\beta\beta$ limit $T_{1/2} > 1.9 \times 10^{25}$~yr while the red points are additionally testable in the near future, i.e. they correspond to $3.0 \times 10^{26} {\rm yr} > T_{1/2} > 1.9 \times 10^{25} $~yr.}
\label{fig:scatter_0nubb}
\end{figure}
In our earlier discussion regarding the $0\nu\beta\beta$ half life, we have fixed the heavy scales of the model as $M_N = M_{\delta^{--}_R}=0.75$~TeV and $M_{W_R}=2.2$~TeV. Instead of fixing these model parameters, we now randomly vary them independently within a range, $M_N = M_{W_R} = M_{\delta^{--}_R} = \text{100}$~GeV - $100$~TeV and the lightest neutrino mass $m_{\rm lightest} = 10^{-4} - 1.0$~eV. The correspondingly generated parameter points are projected on the $M_{W_R} - m_{\rm lightest}$ plane in Fig.~\ref{fig:scatter_0nubb}. The  oscillation parameters are at their best-fit values taken from Table.~\ref{table-osc}. The left and right panels of Fig.~\ref{fig:scatter_0nubb} are in the case of NH and IH light neutrinos, respectively. The green points satisfy the current $0\nu\beta\beta$ half life limit $T_{1/2} > 1.9 \times 10^{25}$~yr while the red points are additionally testable in the near future, i.e. they correspond to $3.0 \times 10^{26} > T_{1/2} > 1.9 \times 10^{25} $~yr. Including the cosmological bound $m_{\rm lightest} \leq 0.077$~eV, we see from Fig.~\ref{fig:scatter_0nubb} that in the case of both NH and IH of light neutrinos, most of the parameter space spanning $ M \lsim \mbox{1\, TeV}$ and  $m_{\rm lightest} \gsim  0.077 {\rm eV})$ is almost ruled out since $M \lsim 1$ TeV will lead to $T_{1/2} \lsim 10^{25}$~yr, which is in contradiction with the current observation from $0\nu\beta\beta$ decay.

\section{Low Energy Lepton Flavour Violating Processes}
\label{sec:LFV}
\begin{figure}[t]
\centering
\includegraphics[clip,width=0.44\textwidth]{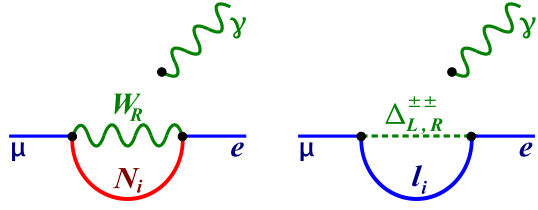}
\caption{Diagrams contributing to $\mu\to e\gamma$ in the LRSM.}
\label{fig:diagrams_LFV_muegamma} 
\end{figure}

\begin{figure}[t]
\centering
\includegraphics[clip,width=0.44\textwidth]{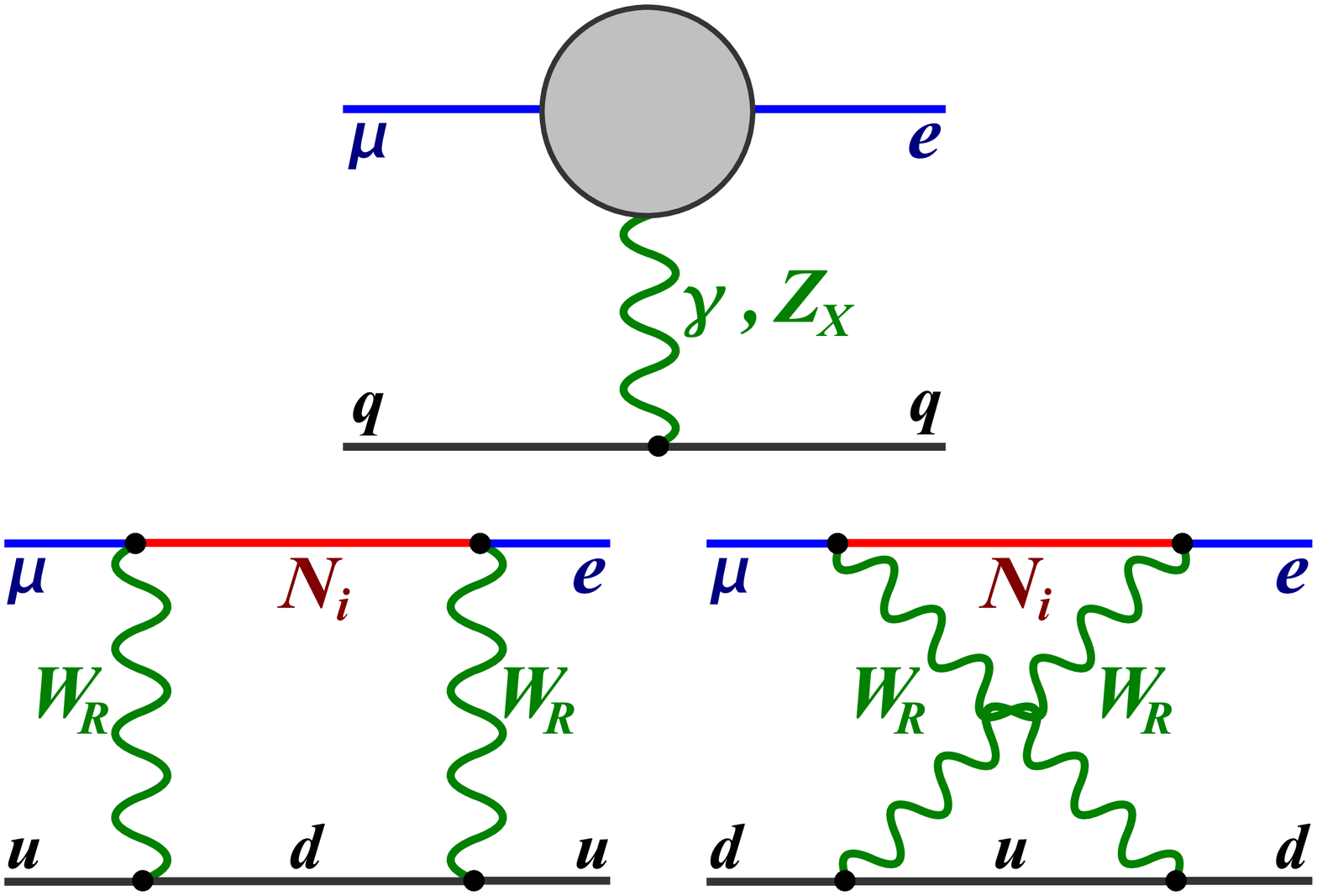}
\includegraphics[clip,width=0.44\textwidth]{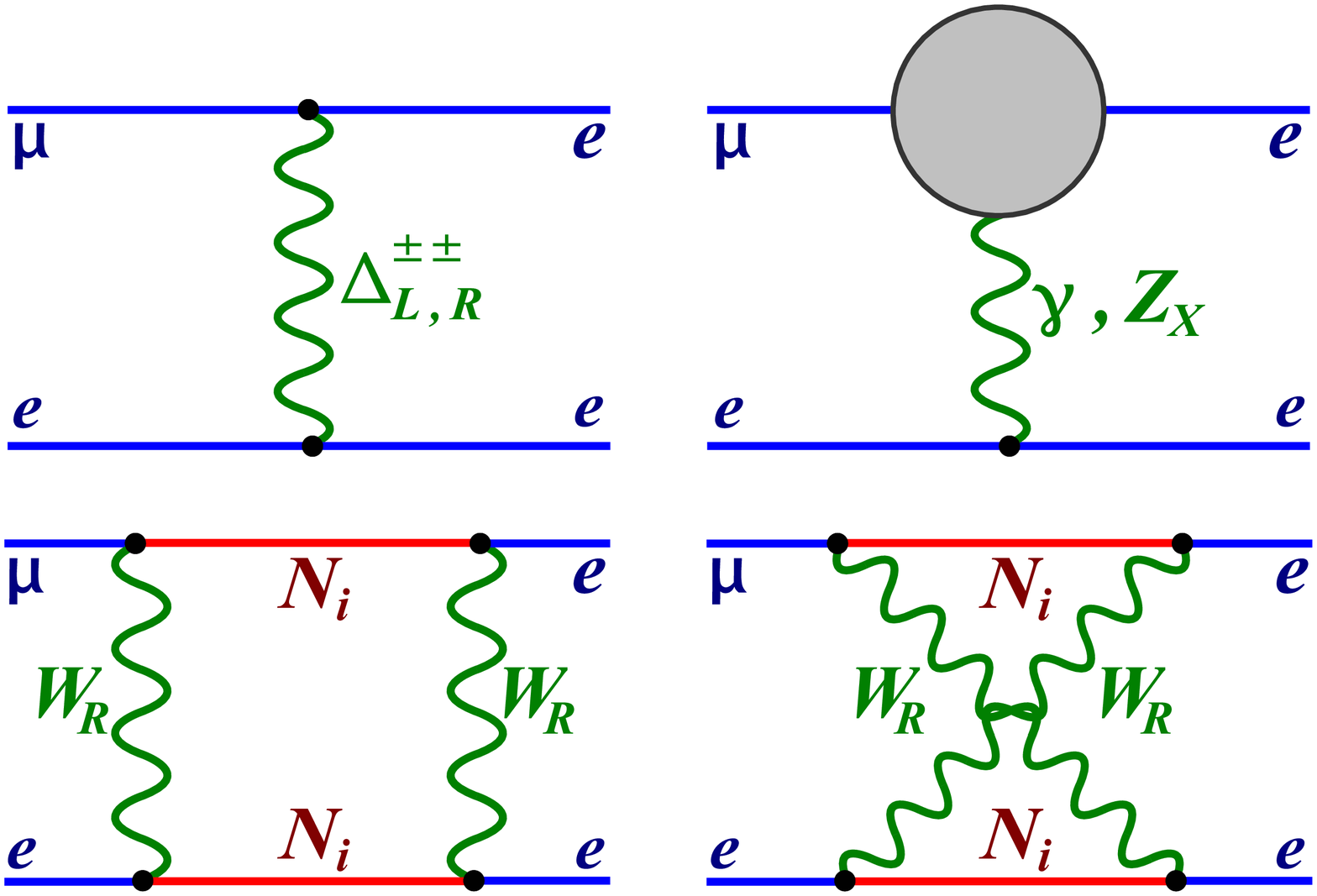}
\caption{Dominant diagrams contributing to $\mu\to e$ conversion in nuclei (left) and $\mu\to eee$ (right) in the Left-Right symmetric model. The grey circle represents the effective $\mu-e-$gauge boson vertex with contributions from Figure~\ref{fig:diagrams_LFV_muegamma}.}
\label{fig:diagrams_LFV_mue} 
\end{figure}

The observation of neutrino oscillations suggests that lepton flavour violation should also take place in other processes. In the given model, the mechanism of light Majorana neutrino mass generation is tightly connected to the phenomenon of charged lepton flavour violation (LFV). Due to the dominance of the Seesaw type II scheme, the mixing among the heavy RH neutrinos is essentially identical as the light neutrino mixing described by the already well-known PMNS matrix $U_\text{PMNS}$. If only light neutrinos were to contribute, LFV is hugely suppressed by the GIM mechanism, i.e. $(\Delta m^2_\nu/m_W^2) \approx 10^{-50}$. The resulting LFV process rates are far below from any experimental sensitivity. On the other hand, sizeable charged lepton flavour violation naturally occurs in the LRSM due to the contributions from the heavy RH neutrinos and Higgs scalars. Because of their sensitivity, we here focus on low energy LFV processes $\mu\to e\gamma$, $\mu\to eee$ and $\mu\to e$~conversion in nuclei, and for example do not consider LFV $\tau$ decays. For a review of LFV and new physics scenarios, see for example \cite{Deppisch:2012vj}. In the LRSM, the processes are described by the diagrams shown in Figures~\ref{fig:diagrams_LFV_muegamma} and \ref{fig:diagrams_LFV_mue}, respectively.

The branching ratios $Br(\mu\to e\gamma)$ and $Br(\mu\to eee)$, as well as the conversion rate $R^N(\mu\to e)$ in a nucleus have been calculated in \cite{Cirigliano:2004mv}, within the context of the LRSM. In general, these processes depend on a large number of parameters, but when assuming that all heavy mass scales have the same order of magnitude, $m_{N_i} \approx m_{W_R} \approx m_{\delta_L^{--}} \approx m_{\delta_R^{--}}$, the result can be greatly simplified. Such a spectrum is naturally expected, as all masses are generated in the breaking of the RH symmetry. The expected process rates are then given by~\cite{Cirigliano:2004mv}
\begin{align}
\label{eq:BrmuegammaSimplified}
	Br(\mu\to e\gamma) &\approx 1.5 \times 10^{-7} |g_{e\mu}|^2 
	\left(\frac{g_R}{g_L}\right)^4
	\left(\frac{1\text{ TeV}}{m_{W_R}}\right)^4, \\
\label{eq:BrmueSimplified}
	R^N(\mu\to e) &\approx X_N \times 10^{-7} |g_{e\mu}|^2
	\left(\frac{g_R}{g_L}\right)^4
	\left(\frac{1\text{ TeV}}{m_{\delta_R^{--}}}\right)^4 
	\alpha \left(\log\frac{m^2_{\delta_R^{--}}}{m^2_{\mu}}\right)^2, \\
\label{eq:BrmueeeSimplified}
	Br(\mu\to eee) &\approx \frac{1}{2}|h_{e\mu}h^*_{ee}|^2 
	\left(\frac{g_R}{g_L}\right)^4
	\left(\frac{m_{W_L}^4}{m^4_{\delta_R^{--}}} +
	\frac{m_{W_L}^4}{m^4_{\delta_L^{--}}}\right).
\end{align}
In \eqref{eq:BrmueSimplified}, $X_\text{(Al,Ti,Au)} \approx (0.8, 1.3 ,1.6)$ is a nucleus-dependent form factor and $g_{e\mu}$, $h_{ij}$ describe the effective lepton-gauge and lepton-Higgs LFV couplings,
\begin{align}
\label{eq:LFVCouplings}
	g_{e\mu} &= 
		\sum_{n=1}^3 V^*_{en} V^{\phantom{\dagger}}_{\mu n}
		\left(\frac{m_{N_n}}{m_{W_R}}\right)^2, \\
	h_{ij} &= 
		\sum_{n=1}^3 V_{in} V_{jn}
		\left(\frac{m_{N_n}}{m_{W_R}}\right), \quad i,j=e,\mu,\tau.
\end{align}
As shown in \cite{Cirigliano:2004mv}, the above formula are valid if the masses generated in breaking the RH symmetry are of the same order, $0.2 \lsim m_i/m_j \lsim 5$ for any pair of $m_{i,j} = m_{N_n}, m_{W_R}, m_{\delta_L^{--}}, m_{\delta_R^{--}}$.

The Equations~(\ref{eq:BrmuegammaSimplified}) - (\ref{eq:BrmueeeSimplified}) immediately allow the following observations: 
(i) Both $Br(\mu\to e\gamma)$ and $R^N(\mu\to e)$ are proportional to the common LFV factor $|g_{e\mu}|^2$. In addition, the ratio of their rates is $R^N(\mu\to e)/Br(\mu\to e\gamma) = \mathcal{O}(1)$, largely independent of the heavy particle spectrum, due to the enhancement of the doubly-charged Higgs boson contributions to $\mu\to e$ conversion. 
(ii) The LFV couplings are generically such that $|g_{e\mu}| \approx |h^*_{ee}h_{e\mu}|$, and thus $Br(\mu\to eee)/R^N(\mu\to e) = \mathcal{O}(300)$ for $m_{\delta_{L,R}^{--}} \approx$ 1 TeV. This can be understand as $\mu \to eee$ is mediated at the tree level in the LRSM, cf. Fig.~\ref{fig:diagrams_LFV_mue}~(right). 
(iii) Due to the Seesaw type II dominance in our model, the LFV couplings \eqref{eq:LFVCouplings} are tightly connected to light neutrino oscillations, $V = U_\text{PMSN}$ and $\Delta M_{ij}^2 = M_{N_i}^2 - M_{N_j}^2 = (M_N / m_{\nu_3})^2 \Delta m_{ij}^2$. 
(iv) The dominant contributions to all three LFV processes are suppressed by a factor $(g_R/g_L)^4$. The above theoretical predictions should be compared with the current experimental upper limits at 90\% C.L.~\cite{Adam:2011ch, Bertl:2006up, Bellgardt:1987du},
\begin{align}
\label{eq:Bexpllgamma}
	Br_{\rm exp}(\mu\to e\gamma) &< 5.7 \cdot 10^{-13}, \nonumber\\
	R^{Au}_{\rm exp}(\mu\to e)   &< 8.0 \cdot 10^{-13}, \\
	Br_{\rm exp}(\mu\to eee)     &< 1.0 \cdot 10^{-12}. \nonumber 
\end{align}
The experimental limits are roughly of the same order, and $Br(\mu\to eee)$ provides the most restrictive bound at the moment. In the near future, the currently running MEG experiment~\cite{Adam:2011ch} aims for the sensitivity
\begin{equation}
\label{eq:BexpllgammaMEG}
	Br_{\rm MEG} (\mu\to e\gamma) \approx 10^{-13},
\end{equation}
and the COMET and Mu2e experiments both plan to reach~\cite{Kutschke:2011ux, Kurup:2011zza}
\begin{equation}
\label{eq:RmueCOMET}
	R^{Al}_{\rm COMET} (\mu\to e) \approx 10^{-16}.
\end{equation}
\begin{figure}[t]
\centering
\includegraphics[clip,width=0.45\textwidth]{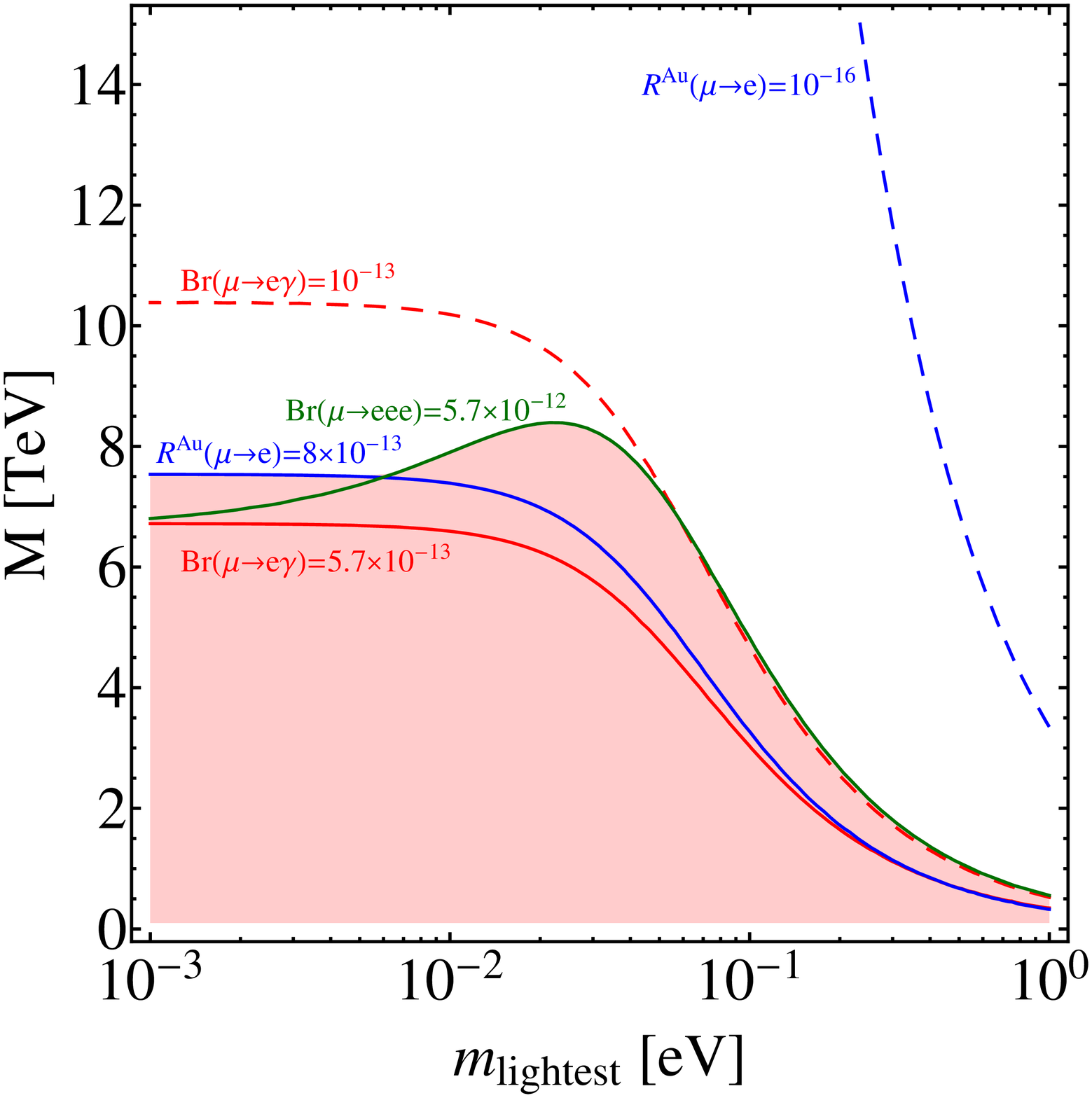}
\includegraphics[clip,width=0.45\textwidth]{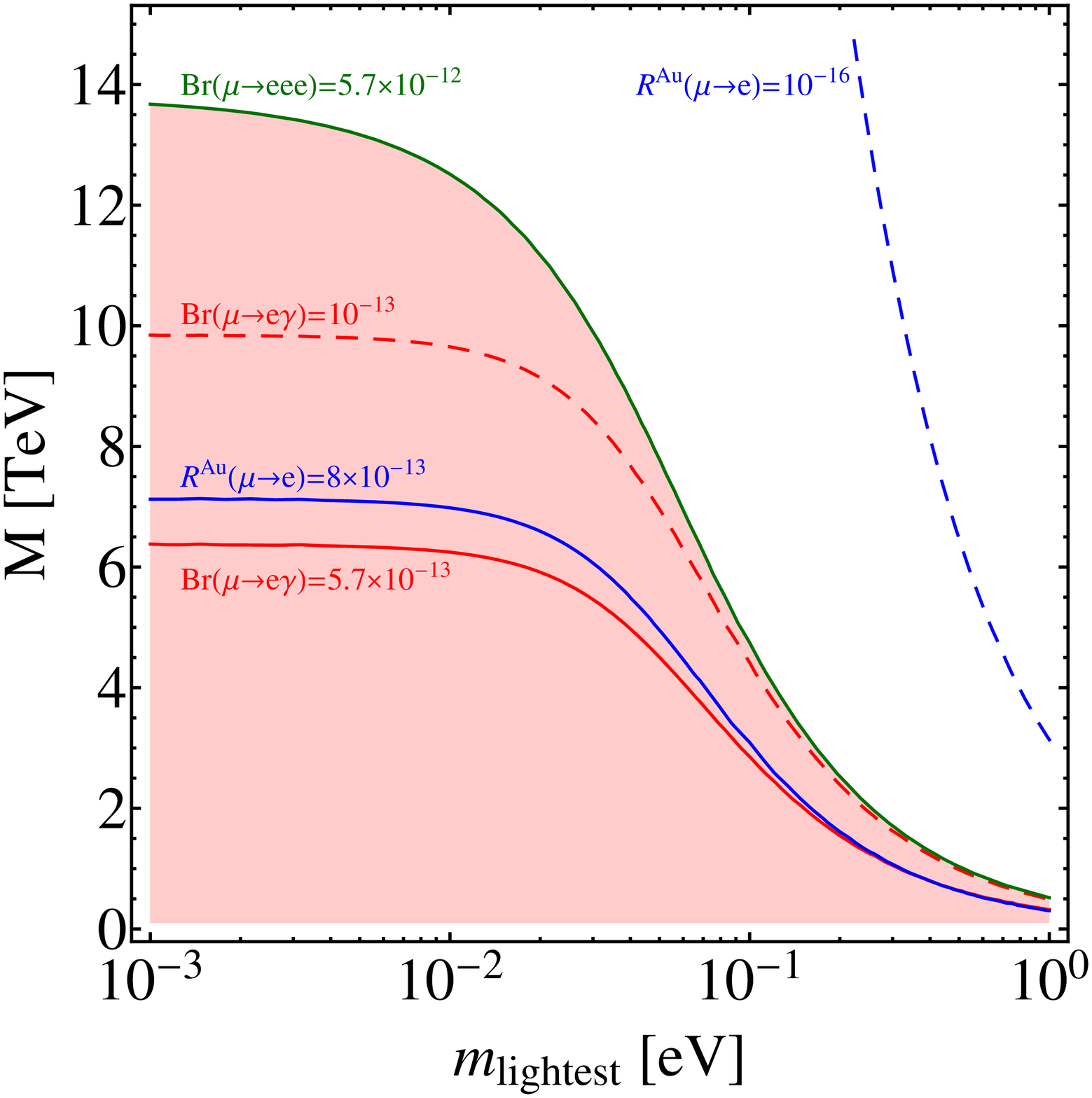}
\caption{Low energy LFV process rates as a function of the lightest neutrino mass $m_\text{lightest}$ and the common heavy mass $M$, in the case of normal (left) and inverse (right) neutrino mass hierarchy. The oscillation parameters are chosen at their best fit values, and the RH gauge coupling $g_R = 0.36$. The solid curves correspond to the respective current limits, whereas the dashed curves give the expected future sensitivities. The shaded area is excluded from current LFV searches.}
\label{fig:lfv_m1_M} 
\end{figure}
The general impact of these limits and sensitivities on the model is shown in Fig.~\ref{fig:lfv_m1_M}. Here we describe the model in terms of the two relevant mass scales: the lightest neutrino mass $m_\text{lightest}$ and the common heavy mass scale $M = M_N, M_{W_R}, M_{\delta_L^{--}}, M_{\delta_R^{--}}$. Putting all masses to the exact same value is a strong simplification, but it allows to easily see the rough dependence. The oscillation parameters are set to their respective best fit values (with a vanishing Dirac $CP$ phase), and the left and right show the case of normal and inverse neutrino mass hierarchy, respectively. The RH gauge coupling is set to its expected minimal value $g_{R} = 0.36$. As expected from the above general considerations, the observable $Br(\mu\to eee)$ currently provides the most stringent constraint on the model parameter space. In any case, it can be seen that LFV searches put a strong constraint on the heavy scale of the model, despite the reduced RH gauge coupling. All LFV 
constraints become weaker for increasing $m_\text{lightest}$ as both the light and heavy neutrino mass spectrum becomes increasingly degenerate resulting in a GIM-like suppression of all LFV effects. In this regard, LFV observables are complementary to $0\nu\beta\beta$ which strongly constrains quasi-degenerate neutrino spectra, cf. Fig.~\ref{fig:scatter_0nubb}. 

\begin{figure}[t!]
\centering
\includegraphics[clip,width=0.49\textwidth]{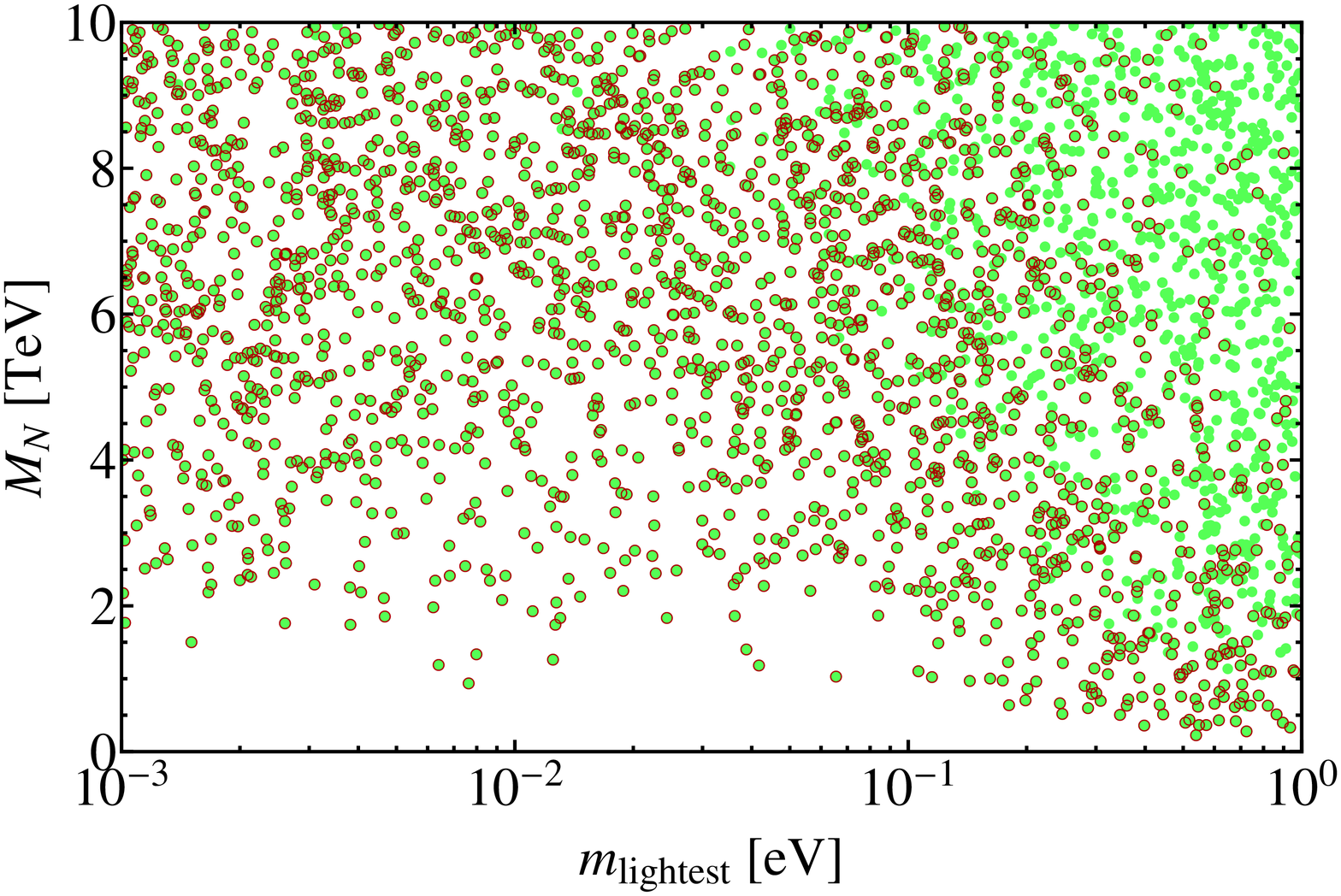}
\includegraphics[clip,width=0.49\textwidth]{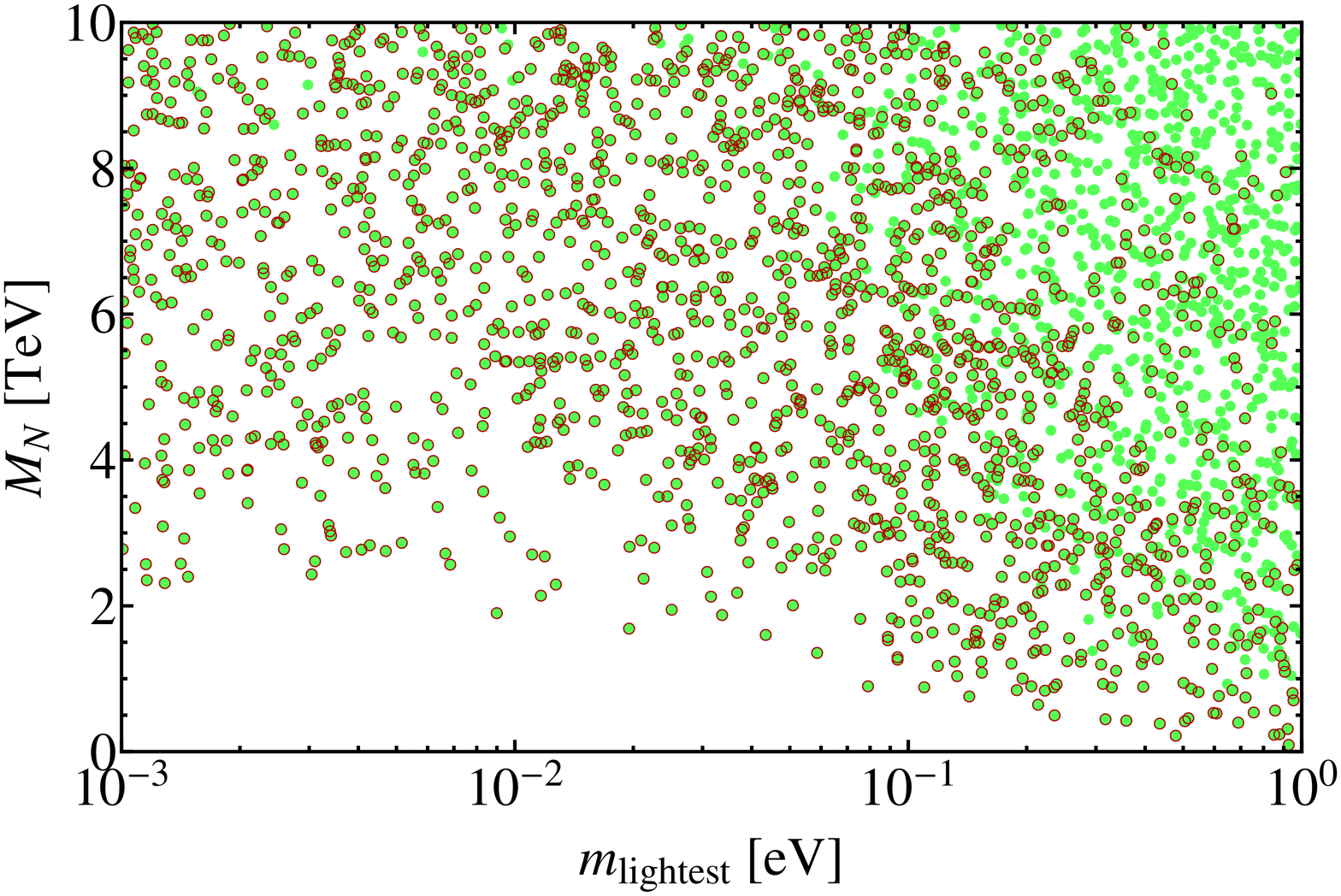}
\caption{Random scatter plot in the parameter plane spanned by $M_N$ and $m_\text{lightest}$. The oscillation parameters are varied according to their experimental errors for normal (left) and inverse (right) light neutrino masses, and the other heavy masses $M_X$ in the range $0.3 M_N < M_X < 3 M_N$. The light green points satisfy all current LFV limits whereas whereas the dark red boxes additionally yield $R^{Al}(\mu\to e) > 10^{-16}$, i.e. can be probed at COMET or Mu2e.}
\label{fig:lfv_m1_M_scatter} 
\end{figure}
Of course, the use of a perfectly commensurate heavy spectrum is an unrealistic simplification. The generalization for a less constrained spectrum is shown in Fig.~\ref{fig:lfv_m1_M_scatter}. Here we scatter the masses $M_X$ of the heavy gauge and Higgs bosons within the range $0.3 M_N < M_X < 3 M_N$, and the oscillation parameters according to their experimental errors around their best fit values. The different colours signify the experimental testability: The light green points are allowed by current LFV searches whereas the dark red boxes additionally yield $R^{Al}(\mu\to e) > 10^{-16}$ i.e. can be probed in the future at COMET or Mu2e. As a result of the additional model freedom, the size of the allowed parameter space increases considerably, and heavy neutrino masses as low as $M_N \approx 2$~TeV are allowed even for hierarchical light neutrino spectra with $m_\text{lightest} \lesssim 10^{-2}$~eV. Nevertheless, the scatter plots in Fig.~\ref{fig:lfv_m1_M_scatter} illustrate that future LFV searches 
such as COMET and Mu2e will be able to probe a large swathe of parameter space.

\section{Collider Signatures}
\label{sec:collider}
\begin{figure}[t!]
\centering
\includegraphics[clip,width=0.46\textwidth]{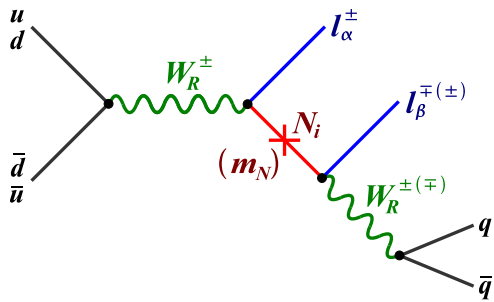}
\caption{Production and decay of a heavy RH neutrino with dilepton signature at hadron colliders.}
\label{fig:diagramsLHC}
\end{figure}

In the LRSM, lepton number and flavour violation can be probed via various processes at the LHC. Generically, the most sensitive process is given by heavy RH neutrino exchange leading to the signal $pp\to W_R \to l_1^\pm l_2^{\pm,\mp} +2$~jets at the LHC \cite{Keung:1983uu, Ho:1990dt, Ferrari:2000sp, Gninenko:2006br, Nemevsek:2011hz}, cf. Fig.~\ref{fig:diagramsLHC}\footnote{Other, possibly non-resonant production mechanisms as well as Triplet Higgs and $Z_R$ production processes can be considered, see for example \cite{Deppisch:2013cya}.}. As can be immediately seen by comparing the diagram with Fig.~\ref{fig:Feyn-type-II-nu}~(middle), the process is essentially the resonant high-energy version of the heavy neutrino contribution to $0\nu\beta\beta$ decay, if two same sign electrons are produced. The potential to discover lepton flavour number violation using this process has analyzed in \cite{AguilarSaavedra:2012fu, Das:2012ii}. We here adapt the analysis of \cite{Das:2012ii} to our specific model and summarize below the salient features. For more details on the LHC process analysis, see \cite{Das:2012ii}.

While we were preparing this manuscript, the CMS collaboration at LHC has reported an updated bound on the mass of the RH charged gauge boson in the LRSM from their analyses of events at a center of mass energy $\sqrt{s} = 8$~TeV with an integrated luminosity of $19.7~{\rm fb}^{-1}$~\cite{cms_excess}. The CMS collaboration looked for the same signature we discuss here, namely two leptons and two jets arising from the $s$-channel production of a $W_R$ boson, which decays through a RH neutrino $N$ as proposed in \cite{Keung:1983uu}, which subsequently decays through an off-shell $W_R$ (other decays of a $W_R$ have been discussed in \cite{Torre_2011}),
\begin{align}
\label{eq:lhcprocess}
	p p \to W_R \to l_1 N \to l_1 l_2 W_R^* \to l_1 l_2 + 2\text{ jets}.
\end{align}
The CMS analysis treats events with two electron and two muons separately but it does not differentiate between lepton charges. Nevertheless, it reports that among the 14 potential signal events seen, only one same sign lepton event was observed.  The analysis does not consider lepton flavour violating signatures with both an electron and a muon. With no significant excess observed in the data, a lower limit on the $W_R$ mass $m_{W_R} < 2.87$ (3.00)~TeV at 90\% CL is reported in the $ee$ ($\mu\mu$) channel, for $M_N = \frac{1}{2} M_{W_R}$. Intriguingly, the data exhibits an excess in the $ee$ channel with a local significance of $2.8\sigma$ for a $W_R$ mass $M_{W_R} \approx 2.1$~TeV. No corresponding excess is seen in the $\mu\mu$ channel. The CMS analysis also reports that it does not see any localized excess in the distribution expected from the decay $N \to l_2 + 2\text{ jets}$.

The CMS analysis compares the experimental result with the theoretically predicted cross section in the minimal LRSM using $g_L = g_R$. Here, the $ee$ excess cannot be understood as the predicted cross section is too large by a factor of $\approx 3-4$. As reported in our dedicated letter \cite{Deppisch:2014qpa}, this issue can be reconciled in LRSMs with $D$ parity breaking that predict a smaller value $g_R \approx 0.6 g_L$. This provides additional motivation to explore the observed excess despite its insufficient significance. Other analyses in the context of LRSMs were performed in~\cite{Heikinheimo:2014tba, Aguilar-Saavedra:2014ola}. The excess has also been discussed in other theoretical contexts~\cite{Bai:2014xba, Dobrescu:2014esa, Allanach:2014lca, Biswas:2014gga,others:2014}.

\begin{figure}[t]
\centering
\includegraphics[width=0.50\linewidth]{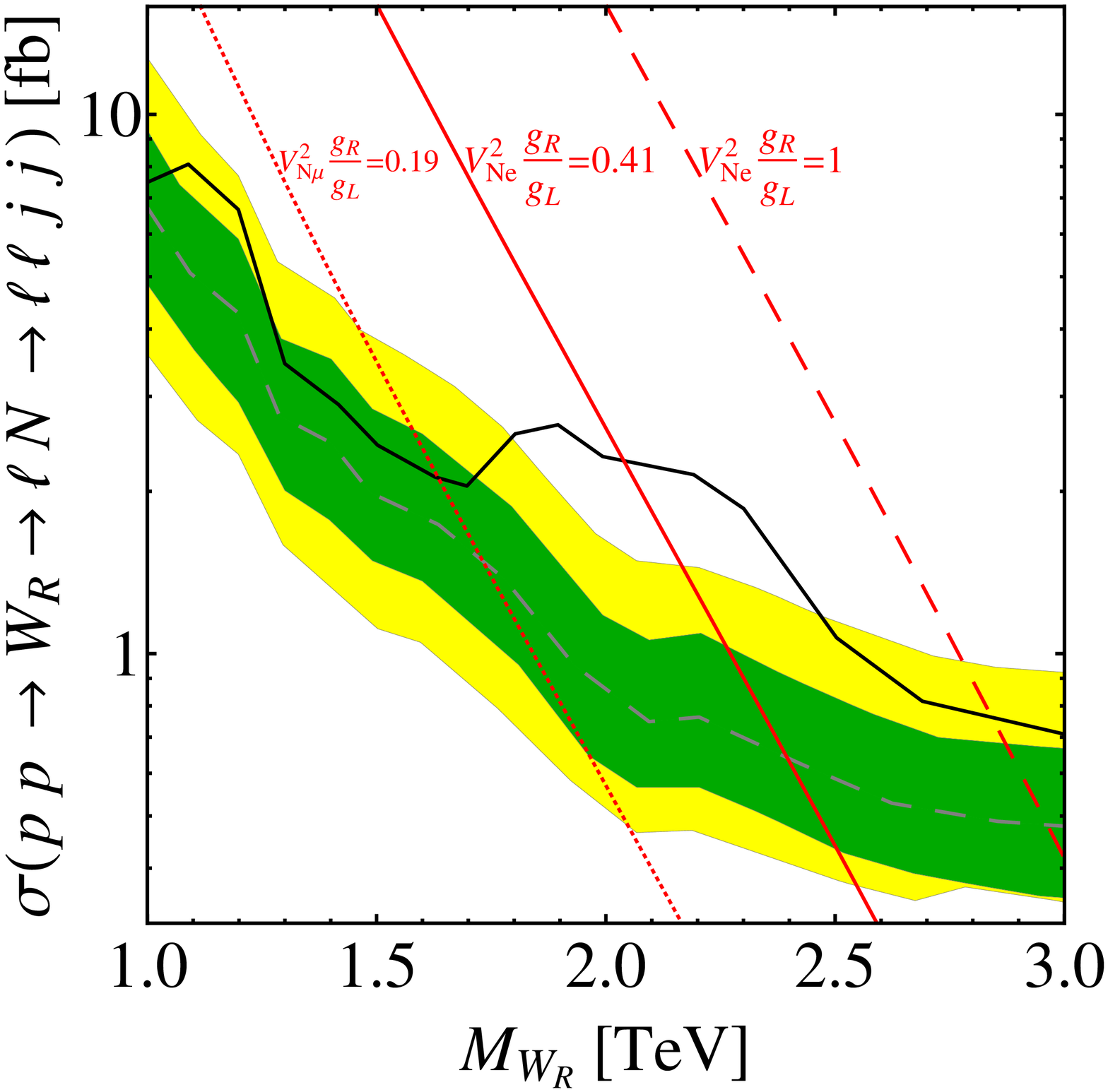}
\includegraphics[clip,width=0.44\textwidth]{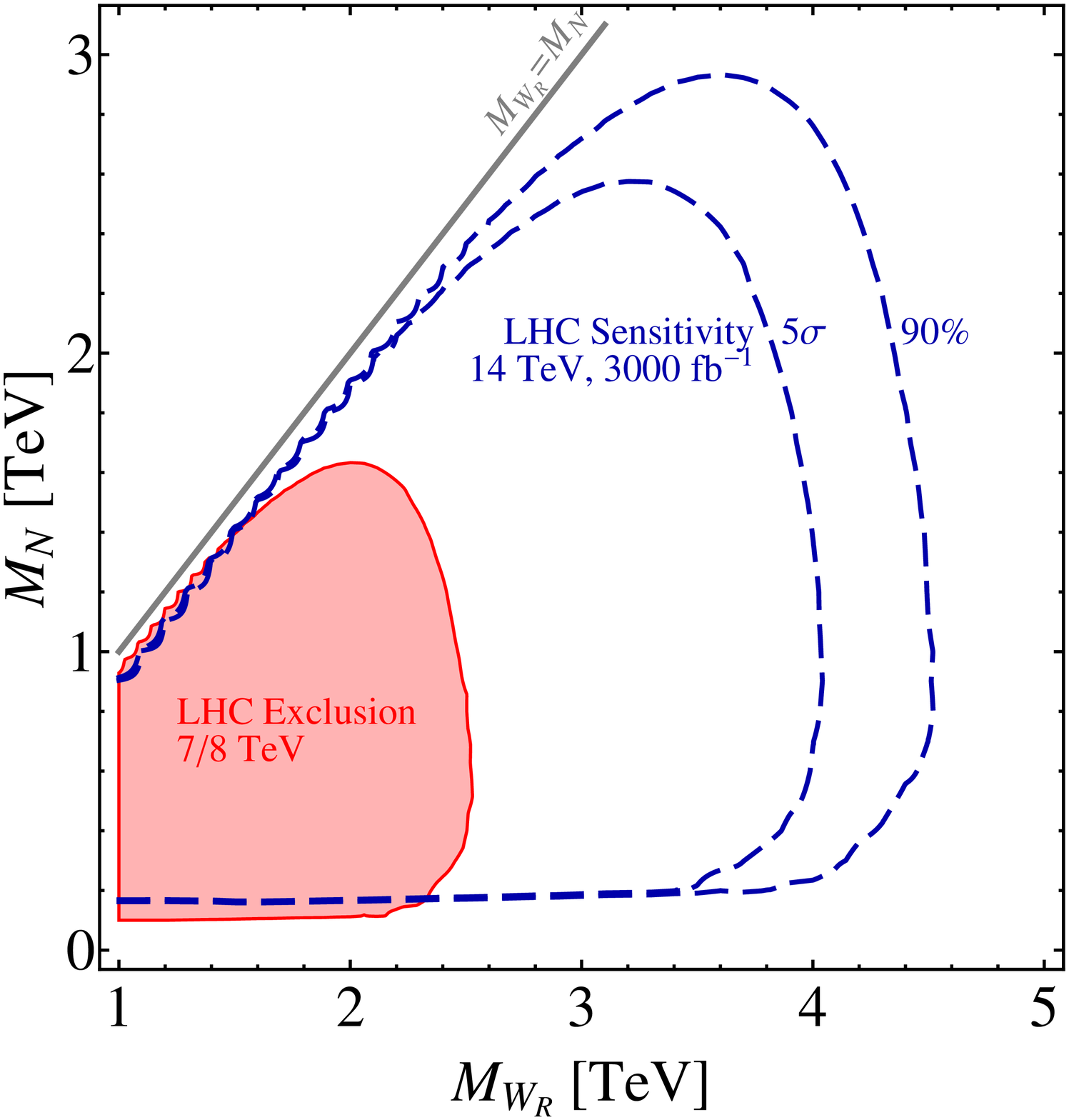}
\caption{Left: Predicted cross sections $\sigma(pp\to eejj)$ (red curves) and experimental exclusion limits as a function of the $W_R$ mass. The dashed red curve correspond to the LRSM case $g_R = g_L$, $V_{Ne} = 1$. The solid (dotted) red curve correspond to the $D$ parity breaking scenario with $g_R = 0.38$ and $|V_{Ne}| = 0.82$ ($|V_{N\mu}| = 0.56$). The observed (solid black curve) and expected (dashed grey curve and green / yellow bands) 95\% exclusion limits are taken from \cite{cms_excess}. Right: Discovery potential of the process $pp\to W_R \to l_1^\pm l_2^{\pm,\mp} +2$~jets at the LHC as a function of the RH $W_R$ boson mass and heavy neutrino mass $M_N$. The red shaded region is excluded by recent LHC searches. The value of the RH gauge coupling is $g_{R}=0.38$, and the neutrino oscillation parameters are chosen at the best fit values.}
\label{fig:sensitivityLHC}
\end{figure}

With the negligible mixing between the heavy and light neutrinos as well as the left and right $W$ bosons, both $W_R$ and $N$ couple only through RH currents. Assuming only one heavy neutrino is light enough to be produced in the process, in a normally ordered hierarchical scenario, the total cross section of the process $pp \to eejj$ can be expressed as
\begin{align}
\label{eq:lhccs}
	\sigma (pp \to eejj) = \sigma (pp \to W_R) 
	\times {\rm Br} (W_R \to e N) 
	\times {\rm Br} (N\to ejj) 
	= |V_{Ne}|^4 \left(\frac{g_R}{g_L}\right)^2 \sigma_\text{CMS}(pp \to eejj),
\end{align}
where $\sigma_\text{CMS}(pp \to eejj)$ is the cross section for $g_L = g_R$ and $V_{Ne} = 1$ (heavy neutrino mixes purely to electrons) as used in the CMS analysis. Instead, in our case we have $g_R / g_L \approx 0.6$ and $|V_{Ne}| \approx 0.85$ for the best fit values of neutrino oscillation parameters. This leads to an overall reduction of the signal cross section by a factor of $|V_{Ne}|^4 (g_R / g_L)^2 \approx 0.16$. This allows the excess to be interpreted as a signal, as shown in Fig.~\ref{fig:sensitivityLHC}~(left) where the calculated process cross section is compared with the CMS result. The dashed red curve gives the predicted cross section as a function of $M_{W_R}$ and $M_N = \frac{1}{2} M_{W_R}$ for $g_R = g_L$ and $V_{Ne} = 1$ (essentially coinciding with the corresponding curve in the CMS analysis) whereas the solid red curve corresponds to $g_R / g_L = 0.6$ and $|V_{Ne}| = 0.82$. The solid black curve is the observed CMS 95\% exclusion whereas the dashed grey curve and green (yellow) bands 
show the expected 95\% exclusion with $1\sigma$ ($2\sigma$) uncertainty, with an excess in the region $1.9 \text{ TeV} \lesssim M_{W_R} \lesssim 2.4$~TeV. The absence of a $\mu\mu$ signal could also be understood as $|V_{N\mu}| \approx 0.56$ for best fit oscillation parameters with a normally ordered neutrino spectrum. The predicted cross section $\sigma(pp \to \mu\mu jj)$ analogous to Eq.~\eqref{eq:lhccs} is also shown in Fig.~\ref{fig:sensitivityLHC}~(left) and clearly would not produce an observable excess.

The non-observation of a signal in previous searches for $W_R \to l N$ at CMS \cite{CMS:2012zv} and ATLAS \cite{ATLAS:2012ak}, which report an exclusion at 95\% confidence level of $M_{W_R} \gtrsim 2.5$~TeV, can of course be easily understood with the suppressed cross section. While apparently incompatible with a signal at $\approx 2.1$~TeV, this limit is also adjusted in our model. With the aforementioned suppression of the cross section, we estimate that the previous LHC limits weaken to $M_{W_R} \gtrsim 2.1$~TeV. A similar argument also apply to other collider limits such as from $W_R \to t \bar b$ decay searches \cite{OtherLhcSearches}.

Despite the apparent strength of our model, we do not consider the excess to be successfully reconciled in our model; firstly, the reported significance of $2.8\sigma$ locally is not high enough, and it is not clear whether the absence of a localized excess in the invariant mass of the heavy neutrino is statistically consistent with the observed excess. Most importantly, though, we have so far omitted the issue of same sign versus opposite sign final state leptons. In our model, the heavy neutrinos are proper Majorana particles, and the same number of positively and negatively charged leptons from its decay are expected. On the other hand, CMS only observed 1 same sign event among the 14 candidate signal events, which would require a strong statistical fluctuation. The model could be enhanced to explain the lack of lepton number violating events by incorporating quasi-Dirac heavy neutrinos, such as present in inverse Seesaw scenarios. While the observation of LNV would provide a clear path understanding 
the physics of neutrino mass generation and would for example have considerable impact on Leptogenesis~\cite{Deppisch:2013jxa, Deppisch:2014hva}, TeV scale models typically predict quasi-Dirac heavy neutrinos to evade unnaturally small Yukawa couplings.

We therefore assume in the following that the excess is only a statistical fluctuation, determine the excluded $W_R$ and $N$ masses in our model and consider the future sensitivity of LHC searches. The potential phenomenology of the process in Fig.~\ref{fig:diagramsLHC} is very rich. As already noted, the final state $e^- e^- + 2$~jets corresponds to lepton number violation directly connected to $0\nu\beta\beta$, whereas leptons with different flavour such as $\mu^+ e^- +2$~jets correspond to lepton flavour violation. First and foremost, though, we are interested in the overall discovery sensitivity, i.e. we sum over all first two generation leptons in the final state, including all combinations of same and opposite sign charges. The discovery sensitivity of future LHC searches with luminosity of 3000$^{-1}$~fb in this case is displayed in Fig.~\ref{fig:sensitivityLHC}~(right). The dashed contours correspond to the $5\sigma$ discovery and 90\% exclusion significance, whereas the red shaded region is already excluded by recent LHC searches~\cite{cms_excess}. In order to determine this region, we assume that the observed CMS sensitivity matches the expected sensitivity, resulting in a limit of $M_{W_R} \gtrsim 2.4$~TeV for $M_N=1/2M_{W_R}$. In our calculation we make the simplifying assumption that the process in Fig.\ref{fig:diagramsLHC} is mediated by a single heavy neutrino, or more precisely, the lightest heavy neutrino. For hierarchical heavy (and light) neutrinos, this will be the dominant contribution. For quasi-degenerate heavy neutrinos, the cross section will increase by a factor of $\approx 3$. Overall, future LHC searches will be able to probe heavy $W_R$ boson and neutrino masses up to $3-4$~TeV in the kinematically allowed regime $m_{W_R} > m_{N}$.

\section{Conclusions}
\label{sec:conclusions}
We have carried out a detailed analysis of the leptonic phenomenology of a class of TeV scale left-right symmetric models with spontaneous breaking of $D$ parity. This includes the low energy aspects of neutrino mass generation within a type-II seesaw mechanism, $0\nu\beta\beta$ decay and lepton flavour violating decays as well as the relevant signatures at the LHC. The main consequence of $D$ parity breaking as compared to minimal models with manifest left-right symmetry is the departure from the left and right gauge coupling equality, $g_R \neq g_L$. Our model emerges from a non-supersymmetric $SO(10)$ GUT scenario with a Pati-Salam symmetry at the highest intermediate scale. The main effect of the model is realized by the reduced value of the RH gauge coupling $g_R \approx 0.6 g_L$, in contrast to most left-right symmetry analyses assuming $g_R = g_L$. 

The reduced RH gauge coupling suppresses all processes mediated by RH currents with various powers of $(g_R / g_L)$ as compared to manifest left-right symmetry. This allows to lower the masses and scales of the model while evading low energy and direct collider limits. We have concentrated on the analyses of the contributions to neutrinoless double beta decay from non-standard effective operators mediated by the heavy states of the model. These processes are suppressed by a factor of $(g_R / g_L)^8 \approx 0.02$ in $W_R-W_R$ mediated channels via the exchange of heavy neutrinos and a heavy RH Higgs triplet. Within the dominant type-II seesaw scenario employed, the different contributions are tightly correlated to the standard light neutrino exchange for a given lepton number symmetry breaking scale, leading to an upper limit on the $0\nu\beta\beta$ half life. 

We have also analysed the predictions for the lepton flavour violating processes $\mu\to e\gamma$, $\mu\to eee$ and $\mu-e$ conversion in nuclei. The dominant contributions to these processes are typically suppressed by $(g_R / g_L)^4$ and LFV searches have therefore a relative advantage over $0\nu\beta\beta$ with respect to the sensitivity to heavy scales. LFV and $0\nu\beta\beta$ searches are also complementary in the sense that while $0\nu\beta\beta$ is especially sensitive to large neutrino masses (quasi-degenerate spectrum), LFV processes are suppressed in this regime, due to a right-handed GIM mechanism. On the other hand, LFV processes are especially sensitive for strong normally or inversely ordered neutrinos, with future experiments such as COMET or Mu2e probing heavy neutrino scales of around 10~TeV.

We have not concentrated on the limits on our model coming from hadronic low energy processes, but we would like to make a few pertinent comments here; The strongest indirect bound on $M_{W_R}$ is due to the $K_L - K_S$ mass difference~\cite{soni_prl, Zhang:2007da, goran_paper_2010, Bertolini:2014sua},
\begin{equation}
	|h_K| \approx \left(\frac{g_R}{g_L}\right)^2 
	              \left(\frac{2.4\text{ TeV}}{M_{W_R}}\right)^2
				< 1.
\end{equation}
This results in the bound $M_{W_R} \gtrsim 2.5$~TeV for manifest left-right symmetry whereas the $(g_R / g_L)^2$ suppression weakens the limit to $M_{W_R} \gtrsim 1.5$~TeV in our case.

Finally, we have also discussed the sensitivity of searches for $W_R$ and $N$ production at the LHC. If the $W_R$ and $N$ only couple via right-handed couplings, as in our model with negligible left-right mixing, the LHC cross section of resonant $W_R$ production scales as $(g_R / g_L)^2$, giving it a relative advantage over low energy LFV and LNV searches. The latest results for this process have been published by CMS recently~\cite{cms_excess} with a reported limit of $M_{W_R} \gtrsim 3$~TeV applying to the manifest left-right symmetry case. The analysed data also includes a local $2.8\sigma$ excess in the cross section of $pp \to e e j j $ for $M_{W_R} \approx 2.0 - 2.5$~TeV. While not sufficiently significant, it is still interesting to speculate whether this excess is generated by new physics beyond the SM. In \cite{Deppisch:2014qpa} we pointed out that the smaller value of $g_R$ in models with $D$ parity breaking is consistent with the observed excess. On the other hand, the predicted cross section is too large in manifest left-right symmetry. We have here elaborated on this observation in our model, in which the right-handed charged current mixing matrix is fully described by the light neutrino oscillation data. This also allows us to understand the absence of an excess in the channel $pp \to \mu\mu j j$ at CMS.

We would like to stress, though, that we do not attempt to explain the excess with our model. We simply consider the excess as a motivation to further explore TeV scale physics models of neutrino mass generation at the LHC. Low energy searches for $0\nu\beta\beta$ and LFV decays together with future LHC searches such as for $W_R$ production with a projected sensitivity of $M_{W_R} \approx 4-5$~TeV (utilizing an integrated luminosity of up to 3000~fb$^{-1}$) in our model will strongly test TeV neutrino mass models.

\begin{acknowledgments}
SP and NS would like to thank the organizers of the Workshop on High Energy Physics and Phenomenology (WHEPP13), held at Puri, Odisha, India during 12-21 December 2013 where this work was initiated. The work of SP is supported by the Department of Science and Technology, Govt. of India under the financial grant SB/S2/HEP-011/2013. The work of NS is partially supported by the Department of Science and Technology, Govt. of India under the financial grant SR/FTP/PS-209/2011. The work of FFD and TEG was supported partly by the London Centre for Terauniverse Studies (LCTS), using funding from the European Research Council via the Advanced Investigator Grant 267352. The work of US is partially supported by the J.C. Bose National Fellowship grant from the Department of Science and Technology, India.
\end{acknowledgments}


\end{document}